\documentclass[11pt]{article}
\usepackage{amsmath}
\usepackage{amsfonts}
\usepackage{amssymb}
\usepackage{graphicx}
\usepackage{multirow}

\title{{\bf {MPEG-2} Prediction Residue Analysis}}
\author{David V\'azquez-Pad\'in$^*$ and Fernando P\'erez-Gonz\'alez\thanks{Signal Theory and Communications Department, University of Vigo, Vigo, Spain (e-mail: dvazquez@gts.uvigo.es; fperez@gts.uvigo.es)}}
\date{}

\begin{document}
\maketitle

\section{Introduction}
\label{sec:intro}

This technical report complements the work in \cite{EUSIPCO}. Based on the use of synthetic signals and autoregressive models to characterize temporal dependencies and inter predictions, here we build a semi-analytic model that explains the evolution of the prediction residue in the MPEG-2 double compression scenario assumed in \cite{EUSIPCO}. As outlined in \cite{EUSIPCO}, this characterization provides valuable insights on the behavior of the Variation of Prediction Footprint (VPF), exploited by different methods (e.g., \cite{VAZQUEZ-12,GVPF-19}), and also shows that the performance of these VPF-based techniques for GOP size estimation and double compression detection depend on the deadzone width of the scalar quantizers used for encoding each type of MacroBlock (MB).

In the following, we first formulate and model the MPEG-2 video double quantization problem in Section~\ref{sec:prob_form_and_mod} to keep this report self-contained. The distinct evolution of the variance of the inter-prediction residue is then analytically characterized in Section~\ref{sec:pred_res_analysis} through the use of the semi-analytic model, and finally, Section~\ref{sec:conclusions} concludes this report and hints at possible new research directions to be explored in the future.

\section{Problem Formulation and Modeling}
\label{sec:prob_form_and_mod}

Let us consider a video double compression scenario where the two encodings are performed with the same MPEG-2 encoder. During the first compression, we assume that the input video sequence is compressed with a constant GOP of length $\text{G}_1$ and a fixed quantization parameter $\text{Q}_1\in\{2,\dots,31\}$. Similarly, the succeeding second compression is conducted with a GOP of length $\text{G}_2$ (different from any integer multiple of submultiple of $\text{G}_1$) and a fixed quantization parameter $\text{Q}_2\in\{2,\dots,31\}$. For the sake of simplicity, we assume that no temporal shift is introduced between both encodings and we discard the use of B-frames, leaving the analysis of bipredictive residues for a future work.

In MPEG-2, the MBs of an I-frame can only be encoded by means of a single intra-coding mode that does not perform any spatial/temporal prediction and is denoted by I-MB. In the case of P-frames, besides the use of I-MBs, two inter-coding modes are available to perform temporal (or motion-compensated) predictions from the last decoded frame: P-MB, which encodes the motion vector and the prediction residue, and S-MB, which efficiently signals those inter-predicted MBs that yield a zero-valued motion vector and null residual data. Accordingly, the set of available coding modes in this case is $\mathcal{C}\triangleq\{\text{I-MB, \text{P-MB}, \text{S-MB}}\}$. To identify the type of encoding a frame has undergone at a particular time index $n$ during the first compression, we define the sets $\mathsf{I}_1$ and  $\mathsf{P}_1$, which respectively contain the time indices of I- and P-frames.

\begin{figure}[t]
\centering
\includegraphics[width=0.99\linewidth]{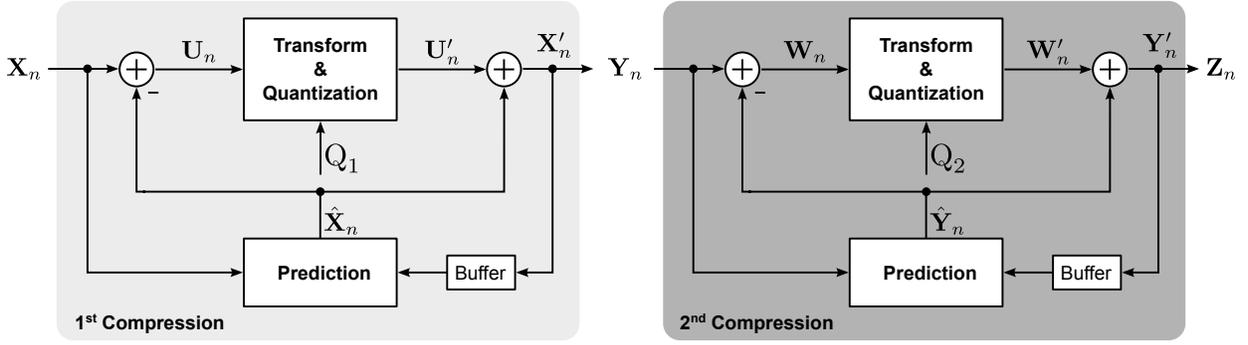}
\caption{Double compression scheme: the left block diagram shows the first compression stage and, correspondingly, the right block depicts the structure of the second compression stage.}
\label{fig:double_comp}
\end{figure}

Under this setting, the block diagram illustrated in Fig.~\ref{fig:double_comp} summarizes the main variables involved in the whole double encoding process. The left scheme in Fig.~\ref{fig:double_comp} models how a given MB at time index $n$, denoted by $\mathbf{X}_n$, is predicted based on a set of previously coded and reconstructed samples stored in a buffer.\footnote{The position indices showing the location of the MB $\mathbf{X}_n$ within the frame are omitted for the sake of clarity.} Depending on the coding mode $\text{c}\in\mathcal{C}$ selected by the encoder, the prediction $\hat{\mathbf{X}}_n$ is computed as
\begin{equation}
\hat{\mathbf{X}}_n=\begin{cases}
0, & \text{if } \text{c}=\text{I-MB}\\
\mathbf{X}_{n-1}^{\prime}(\mathbf{m}), & \text{otherwise}
\end{cases},
\label{eq:hat_X_n}
\end{equation}
where $\mathbf{X}_{n-1}^{\prime}(\mathbf{m})$ denotes the MB extracted from the reference frame (previously decoded at time index $n-1$) with the relative displacement that the motion vector $\mathbf{m}$ points out. The first case in \eqref{eq:hat_X_n} reflects that no prediction is used for I-MBs, while the second case is valid for representing the motion-compensated prediction of P-MBs and also that of S-MBs, provided that $\mathbf{m}=(0,0)$.

After the prediction, a residue is obtained as $\mathbf{U}_n=\mathbf{X}_n-\hat{\mathbf{X}}_n$, which is later transformed applying the Discrete Cosine Transform (DCT) on an $8\times8$ block-basis. In the DCT domain, each $(i,j)$-th coefficient with $i,j\in\{0,\dots,7\}$ is quantized with a distinct quantization step size and a configurable deadzone width. The quantization step size is modified by two mechanisms: a weighting matrix $\mathbf{S}$ to improve the perceptual quality of the encoded videos and a scale factor (controlled by the quantization parameter $\text{Q}_1$) that globally adapts the size of each quantization step, having
\begin{equation} 
\Delta_1(i,j)\triangleq\text{Q}_1\frac{S_{i,j}}{8},\quad\forall i,j\in\{0,\dots,7\},
\label{eq:quant_step}
\end{equation}
where $S_{i,j}$ represents the $(i,j)$-th element of the matrix $\mathbf{S}$. MPEG-2 supports the use of different quantization weighting matrices for intra- and inter-coding modes. As an example, the MPEG-2 encoder implementation from the FFmpeg library \cite{FFMPEG} uses by default the following weighting matrix $\mathbf{S}^{\text{I}}$ for intra-coding modes
\begin{equation*}
\mathbf{S}^{\text{I}}=\left(\begin{array}{cccccccc}
    8 & 16 & 19 & 22 & 26 & 27 & 29 & 34\\
    16 & 16 & 22 & 24 & 27 & 29 & 34 & 37\\
    19 & 22 & 26 & 27 & 29 & 34 & 34 & 38\\
    22 & 22 & 26 & 27 & 29 & 34 & 37 & 40\\
    22 & 26 & 27 & 29 & 32 & 35 & 40 & 48\\
    26 & 27 & 29 & 32 & 35 & 40 & 48 & 58\\
    26 & 27 & 29 & 34 & 38 & 46 & 56 & 69\\
    27 & 29 & 35 & 38 & 46 & 56 & 69 & 83
\end{array}\right),
\end{equation*}
whereas for inter-coding modes a weighting matrix $\mathbf{S}^{\text{P}}$ is adopted with $S_{i,j}^{\text{P}}=16, \forall i,j\in\{0,\dots,7\}$. Regarding the quantizer deadzone, its width is defined as a function of the applied quantization step, i.e., 
\begin{equation}
w_1(i,j)\triangleq\alpha\Delta_1(i,j),
\label{eq:deadzone}
\end{equation}
where $\alpha\in[1,2]$ is the parameter that allows the control of the deadzone width. The use of a wider deadzone generally leads to a lower bitrate because more transform coefficients are quantized to zero, but in contrast a higher degree of distortion is introduced. Hence, in practice, the use of wider deadzones is recommended for small magnitude signals, such as the ones resulting from the use of inter-coding modes, whereas tighter deadzones are more convenient for intra-coding modes to retain more details in key reference frames. For instance, the MPEG-2 encoder in \cite{FFMPEG} uses by default different values of $\alpha$ for intra- and inter-coding modes, namely: $\alpha_{\text{I}}=\frac{5}{4}$ and $\alpha_{\text{P}}=2$, respectively. Further in this work, the impact on the variance of the prediction residue in the second compression stage will be analyzed for $\alpha_{\text{P}}=2$ and different values of $\alpha_{\text{I}}$, keeping them constant across both compressions.

Now, using \eqref{eq:quant_step} and \eqref{eq:deadzone}, the quantization of a given AC coefficient $u$ from the DCT of $\mathbf{U}_n$ can be written (omitting the position indices) as
\begin{equation}
u_q\triangleq\begin{cases}
\text{sgn}(u)\left\lfloor\frac{|u|+\Delta_1^{\text{I}}\left(1-\frac{\alpha_{\text{I}}}{2}\right)}{\Delta_1^{\text{I}}}\right\rfloor, & \text{if } \text{c}=\text{I-MB}\\
\\
\text{sgn}(u)\left\lfloor\frac{|u|+\Delta_1^{\text{P}}\left(1-\frac{\alpha_{\text{P}}}{2}\right)}{\Delta_1^{\text{P}}}\right\rfloor, & \text{otherwise}
\end{cases},
\label{eq:u_q}
\end{equation}
where $|\cdot|$ is the absolute value operator, $\lfloor\cdot\rfloor$ denotes the floor function, and $\text{sgn}(\cdot)$ represents the sign function which returns $-1$ for a negative number, $0$ for the number zero, and $+1$ for a positive number. The notation $\Delta_1^{\text{I}}$ and $\Delta_1^{\text{P}}$ has been used to remark that different quantization steps can be employed in each coding mode, depending on which quantization weighting matrix $\mathbf{S}^{\text{I}}$ or $\mathbf{S}^{\text{P}}$ is used, respectively.

According to the MPEG-2 standard, the de-quantized version $u^\prime$ of a quantized AC coefficient $u_q$ is given by
\begin{equation}
u^\prime = \begin{cases}
\text{sgn}\left(u_q\right)\left\lfloor\Delta_1^{\text{I}} |u_q|\right\rfloor, & \text{if c=I-MB}\\
\\
\text{sgn}\left(u_q\right)\left\lfloor\Delta_1^{\text{P}} |u_q| + \frac{\Delta_1^{\text{P}}}{2}\right\rfloor, & \text{otherwise}
\end{cases},
\label{eq:u_prime}
\end{equation}
where a specific process is followed depending on the applied coding mode, i.e., the reconstructed values for I-MBs are distributed on an equally spaced grid (determined by $\Delta_1^{\text{I}}$), while for inter-coded MBs the first nonzero reconstructed value is shifted a distance of $\frac{3}{2}\Delta_1^{\text{P}}$ from zero. The use of different rules for de-quantization contributes to improving the coding performance, and as we will further see in Section~\ref{sec:pred_res_analysis}, it also affects the variance of the prediction residue in the second compression stage. As a last step in the reconstruction process, the samples in the pixel domain $\mathbf{X}^\prime_n$ are recovered by adding back the de-quantized and inverse transformed samples $\mathbf{U}^\prime_n$ to the prediction $\hat{\mathbf{X}}_n$, such that $\mathbf{X}^\prime_n=\mathbf{U}^\prime_n+\hat{\mathbf{X}}_n$. 

The above description straightforwardly extends to the second compression block on the right of Fig.~\ref{fig:double_comp}: the source and predicted samples are denoted by $\mathbf{Y}_n$ and $\hat{\mathbf{Y}}_n$, respectively, the residue signal by $\mathbf{W}_n$ and its reconstructed version by $\mathbf{W}^\prime_n$; in this case, the quantization parameter is denoted by $\text{Q}_2$, the quantization steps by $\Delta_2^{\text{I}}$ and $\Delta_2^{\text{P}}$, and the recovered samples are accordingly represented by $\mathbf{Y}^\prime_n$.

To make this problem analytically tractable, the upcoming analysis will be theoretically supported by focusing on a single DCT coefficient (i.e., the one at position $(1,0)$) and assuming that the input samples from $\mathbf{X}_n$ in the DCT domain follow a Laplacian distribution with mean $\mu_{X}$ and variance $\sigma_X^2$ (the rationale behind the use of the Laplacian model is provided in \cite{LAM-00}). Certainly, this model is too simplistic and lossy, but the theoretical conclusions derived from it will be applicable on a large set of real video sequences.

\section{Prediction Residue Analysis}
\label{sec:pred_res_analysis}

The use of de-synchronized GOPs in a double encoding scheme causes the VPF effect unveiled in \cite{VAZQUEZ-12}, which leads to periodic changes in the distribution of certain MB types in double compressed videos, specifically, at P-frames that were originally encoded as I-frames. In view of the straight connection between the presence of the VPF and the MB type selection process implemented by the encoder, we focus on the nowadays most common strategy for MB coding-mode selection, which is based on Lagrangian optimization \cite{SULLIVAN-98} and consists in solving the following minimization problem
\begin{equation}
\text{MB type} = \arg \min_{\text{c}\in\mathcal{C}} D(\mathbf{Y}_n,\mathbf{Y}^\prime_n)+\lambda_{\text{c}}R(\mathbf{Y}^\prime_n),
\label{eq:lag-mode-decision}
\end{equation}
where $\lambda_{\text{c}}$ denotes the Lagrange multiplier of the coding mode $\text{c}\!\in\!\mathcal{C}$, the distortion $D(\mathbf{Y}_n,\mathbf{Y}^\prime_n)$ is the Sum of Squared Differences (SSD) between the reconstructed block $\mathbf{Y}^\prime_n$ and its source $\mathbf{Y}_n$, and the rate $R(\mathbf{Y}^\prime_n)$ measures the number of required bits to reconstruct $\mathbf{Y}^\prime_n$. From \eqref{eq:lag-mode-decision} and since $$D(\mathbf{Y}_n,\mathbf{Y}^\prime_n)\triangleq\|\mathbf{Y}_n-\mathbf{Y}_n^\prime\|_2^2=\|\mathbf{W}_n+\hat{\mathbf{Y}}_n-(\mathbf{W}^\prime_n+\hat{\mathbf{Y}}_n)\|_2^2=\|\mathbf{W}_n-\mathbf{W}^\prime_n\|_2^2,$$ we know that the selection of a particular MB type depends on the variance of the prediction residue. So, to predict the strength of the VPF on those P-frames originally encoded as I-frames, we need to analyze the evolution of the difference of the variance $\text{Var}\left(\mathbf{W}_n\right)$ under $n\!\in\!\mathsf{I}_1$ and $n\!\in\!\mathsf{P}_1$, i.e.,
\begin{equation}
\text{Var}\left(\mathbf{W}_n\right)|_{n\in\mathsf{I}_1}-\text{Var}\left(\mathbf{W}_n\right)|_{n\in\mathsf{P}_1},
\label{eq:W_n}
\end{equation}
where a larger difference value yields a stronger VPF. In the following, we carry out a comprehensive analysis of \eqref{eq:W_n}, particularizing the cases in which solely predictions of type intra (Sect.~\ref{subsec:intra}) or inter (Sect.~\ref{subsec:inter}) are used during the second compression.

Let us characterize the prediction residue $\mathbf{W}_n=\mathbf{Y}_n-\hat{\mathbf{Y}}_n$ by first describing the input signal $\mathbf{Y}_n$, which can be expressed as
\begin{align*}
\mathbf{Y}_n &= \mathbf{X}^\prime_n\\
&=\mathbf{U}^\prime_n+\hat{\mathbf{X}}_n\\
&=\mathbf{X}_n+\left(\mathbf{U}^\prime_n-\mathbf{U}_n\right),
\end{align*}
where the relation $\hat{\mathbf{X}}_n=\mathbf{X}_n-\mathbf{U}_n$ has been used in the last step. The above equation shows that the input signal at the second compression stage can be seen as the source signal $\mathbf{X}_n$ with an added quantization error $(\mathbf{U}_n^\prime-\mathbf{U}_n)$ that depends on the selected type of frame (and the correspondingly applied coding modes) during the first compression. 

When $n\in\mathsf{I}_1$, only I-MBs can be used to encode $\mathbf{X}_n$, which implies that $\mathbf{U}^\prime_n-\mathbf{U}_n=\mathbf{X}^\prime_n-\mathbf{X}_n$, since from \eqref{eq:hat_X_n} we have that $\mathbf{U}_n=\mathbf{X}_n$ and $\mathbf{U}_n^\prime=\mathbf{X}_n^\prime$. This particular quantization error is denoted by $\mathbf{E}_n^{\text{I}_1}\triangleq\mathbf{X}^\prime_n-\mathbf{X}_n$. On the other hand, when $n\in\mathsf{P}_1$, we have that $\mathbf{E}_n^{\text{P}_1}\triangleq\mathbf{U}^\prime_n-\mathbf{U}_n$. In consequence, the input signal can be rewritten as
\begin{equation}
\mathbf{Y}_n = \begin{cases}
\mathbf{X}_n+\mathbf{E}_n^{\text{I}_1}, & \text{if } n\in\mathsf{I}_1\\
\mathbf{X}_n+\mathbf{E}_n^{\text{P}_1}, & \text{if } n\in\mathsf{P}_1\\
\end{cases}.
\label{eq:Y_n}
\end{equation}
The subsequent sections separately describe the prediction $\hat{\mathbf{Y}}_n$ as a function of the two coding modes, i.e., intra or inter, that can be applied under the second compression.

\subsection{Intra-prediction residue analysis}
\label{subsec:intra}

The use of an intra-coding mode during the second compression yields $\hat{\mathbf{Y}}_n=0$, such that $\mathbf{W}_n=\mathbf{Y}_n$. Hence, from \eqref{eq:Y_n}, the variance of the resulting prediction residue can be expressed as
\begin{equation}
\text{Var}\left(\mathbf{W}_n\right) = \begin{cases}
\text{Var}\left(\mathbf{X}_n\right)+\text{Var}\left(\mathbf{E}_n^{\text{I}_1}\right), & \text{if } n\in\mathsf{I}_1\\
\text{Var}\left(\mathbf{X}_n\right)+\text{Var}\left(\mathbf{E}_n^{\text{P}_1}\right), & \text{if } n\in\mathsf{P}_1
\end{cases},
\label{eq:var_Wn_intra}
\end{equation}
where we assume that the quantization errors $\mathbf{E}_n^{\text{I}_1}$ and $\mathbf{E}_n^{\text{P}_1}$ have negligible correlation with the source signal $\mathbf{X}_n$. This assumption typically holds whenever the probability density function (pdf) of the source signal is smooth and its variance is much larger than the employed quantization step sizes, which is generally the case in practice. By inserting the relationship \eqref{eq:var_Wn_intra} in \eqref{eq:W_n}, the strength of the VPF can be evaluated in this case by means of 
\begin{equation*}
\text{Var}\left(\mathbf{E}_n^{\text{I}_1}\right)-\text{Var}\left(\mathbf{E}_n^{\text{P}_1}\right).
\end{equation*}
The variance of both quantization errors can be analytically described conforming to the model discussed in Section~\ref{sec:prob_form_and_mod}, as follows:

\begin{enumerate}
\item $\text{Var}\left(\mathbf{E}_n^{\text{I}_1}\right)$: given the definition of $\mathbf{E}_n^{\text{I}_1}$, its variance is proportional (except for a constant normalization factor) to the distortion between the reconstructed samples $\mathbf{X}_n^\prime$ and their source $\mathbf{X}_n$ when the SSD measure is considered, thus having
\begin{equation*}
\text{Var}\left(\mathbf{E}_n^{\text{I}_1}\right)\propto D\left(\mathbf{X}_n,\mathbf{X}_n^\prime\right)\triangleq\|\mathbf{X}_n-\mathbf{X}_n^\prime\|_2^2.
\end{equation*}
Since the DCT adopted in MPEG-2 has orthogonal basis (discarding rounding effects in the DCT calculation), the above distortion can be directly computed in the transformed domain as
\begin{equation*}
D\left(\mathbf{X}_n,\mathbf{X}_n^\prime\right)=\sum_{i=0}^7\sum_{j=0}^7D_X(i,j),
\end{equation*}
where $D_X(i,j)$ represents the distortion of the $(i,j)$-th DCT coefficient. In line with the quantization model introduced in Section~\ref{sec:prob_form_and_mod}, the values of $D_X(i,j)$ (except for the DC coefficient) are given by 
\begin{align}
D_X(i,j)=&\int_{-\frac{\alpha_{\text{I}}}{2}\Delta_1^{\text{I}}}^{\frac{\alpha_{\text{I}}}{2}\Delta_1^{\text{I}}}x^2f_X(x)~dx\nonumber\\
&+ 2\sum_{k=1}^{\infty}\int_{(k-1+\frac{\alpha_{\text{I}}}{2})\Delta_1^{\text{I}}}^{(k+\frac{\alpha_{\text{I}}}{2})\Delta_1^{\text{I}}}\left(x-\left\lfloor k\Delta_1^{\text{I}}\right\rfloor\right)^2f_X(x)~dx,
\label{eq:D_X}
\end{align}
where $f_X(x)$ denotes the pdf of the corresponding AC coefficient which, as discussed at the end of Section~\ref{sec:prob_form_and_mod}, can be modeled as a Laplacian distribution with mean $\mu_{X}$ and variance $\sigma_X^2$.

From the above equation we observe that for a fixed value of $\sigma_X^2$, the evolution of $\text{Var}\left(\mathbf{E}_n^{\text{I}_1}\right)$ mostly depends on the deadzone width (configured through the parameter $\alpha_{\text{I}}$) and on the quantization step $\Delta_{1}^{\text{I}}$ (controlled by the quantization parameter $\text{Q}_1$ as in \eqref{eq:quant_step}). More specifically, we can state that larger values of $\alpha_{\text{I}}$ (or, equivalently, wider deadzones) yield larger values of $\text{Var}\left(\mathbf{E}_n^{\text{I}_1}\right)$, which also increases with $\text{Q}_1$, since larger values of $\text{Q}_1$ produce coarser quantization steps $\Delta_{1}^{\text{I}}$.

\item $\text{Var}\left(\mathbf{E}_n^{\text{P}_1}\right)$: the variance of the quantization error $\mathbf{E}_n^{\text{P}_1}$ satisfies the following relation
\begin{align*}
\text{Var}\left(\mathbf{E}_n^{\text{P}_1}\right)&\propto\text{p}(\text{I-MB})D(\mathbf{X}_n,\mathbf{X}_n^\prime)+\text{p}(\text{P-MB})D(\mathbf{U}_n,\mathbf{U}_n^\prime)\\
&\quad+\text{p}(\text{S-MB})D(\mathbf{X}_n,\mathbf{X}_{n\!-\!1}^\prime),
\end{align*}
where $\text{p}(\text{c})$ denotes the probability of using the coding mode $\text{c}\!\in\!\mathcal{C}$ per frame. Regarding the distortion terms, both $D(\mathbf{X}_n,\mathbf{X}_n^\prime)$ and $D(\mathbf{X}_n,\mathbf{X}_{n-1}^\prime)$ can be computed as in the previous case through \eqref{eq:D_X},\footnote{Note that for $D(\mathbf{X}_n,\mathbf{X}_{n\!-\!1}^\prime)$, only an approximation would be obtained through \eqref{eq:D_X}, but still valid in practice since $\mathbf{X}_n\approx\mathbf{X}_{n\!-\!1}$ for S-MBs.} whereas $D(\mathbf{U}_n,\mathbf{U}_n^\prime)$ can be obtained by accumulating the distortion of each $(i,j)$-th DCT coefficient, i.e., $D(\mathbf{U}_n,\mathbf{U}_n^\prime)=\sum_{i=0}^7\sum_{j=0}^7D_U(i,j)$, where $D_U(i,j)$ is given by
\begin{align*}
D_U(i,j)=&\int_{-\frac{\alpha_{\text{P}}}{2}\Delta_1^{\text{P}}}^{\frac{\alpha_{\text{P}}}{2}\Delta_1^{\text{P}}}u^2f_U(u)~du\\
&+ 2\sum_{k=1}^{\infty}\int_{(k-1+\frac{\alpha_{\text{P}}}{2})\Delta_1^{\text{P}}}^{(k+\frac{\alpha_{\text{P}}}{2})\Delta_1^{\text{P}}}\left(u\!-\!\left\lfloor k\Delta_1^{\text{P}}+\frac{\Delta_1^{\text{P}}}{2}\right\rfloor\right)^2f_U(u)~du.
\end{align*}
In the above equation, $f_U(u)$ denotes the pdf of the $(i,j)$-th DCT coefficient which can be modeled as a Laplacian distribution with zero mean and variance $\sigma_U^2$ \cite{BELLIFEMINE-92}. 

Similarly to the previous case, $\text{Var}\left(\mathbf{E}_n^{\text{P}_1}\right)$ also monotonically increases with $\text{Q}_1$, although now the effect of the deadzone width is a function of both parameters $\alpha_{\text{I}}$ and $\alpha_{\text{P}}$. The influence of each deadzone width will depend on the type of scene to be encoded, since its content will finally rule the probability of using each type of MB. For instance, in static scenes, it is common to have $\text{p}(\text{S-MB})\gg\text{p}(\text{P-MB})+\text{p}(\text{I-MB})$, so $\text{Var}\left(\mathbf{E}_n^{\text{P}_1}\right)$ will increase with $\text{Q}_1$ and $\alpha_{\text{I}}$, and will not be excessively affected by $\alpha_{\text{P}}$, whereas in dynamic scenes, where almost all the coded MBs have a non-zero motion vector with $\mathbf{U}_n^\prime\neq 0$, such that $\text{p}(\text{P-MB})\gg\text{p}(\text{S-MB})+\text{p}(\text{I-MB})$, the evolution of $\text{Var}\left(\mathbf{E}_n^{\text{P}_1}\right)$ will be mostly governed by $\alpha_{\text{P}}$ instead of $\alpha_{\text{I}}$. 
\end{enumerate}

\begin{figure}[t]
\centering
\begin{minipage}{0.32\linewidth}
\centering
\includegraphics[width=\linewidth]{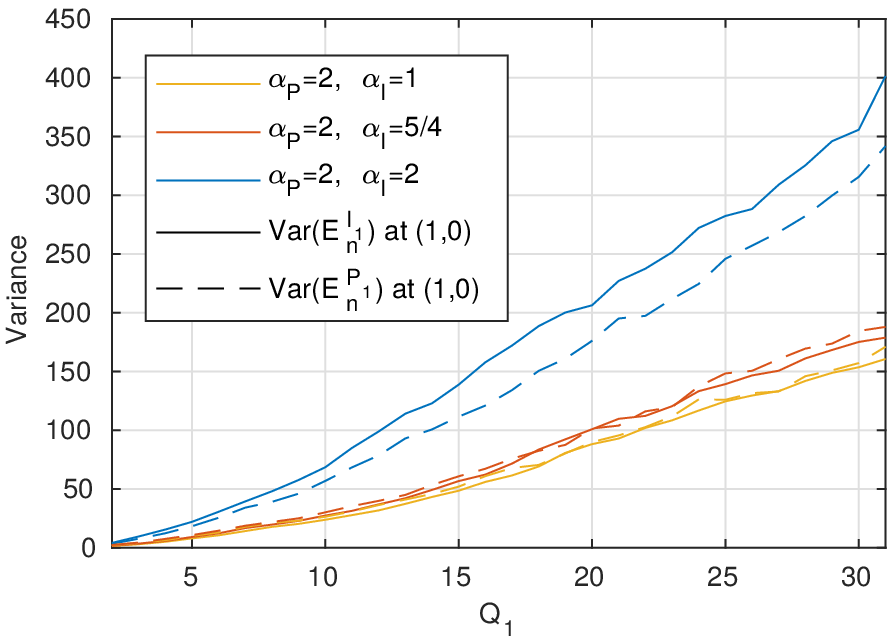}
\centerline{(a)}
\end{minipage}
\hfill
\begin{minipage}{0.32\linewidth}
\centering
\includegraphics[width=\linewidth]{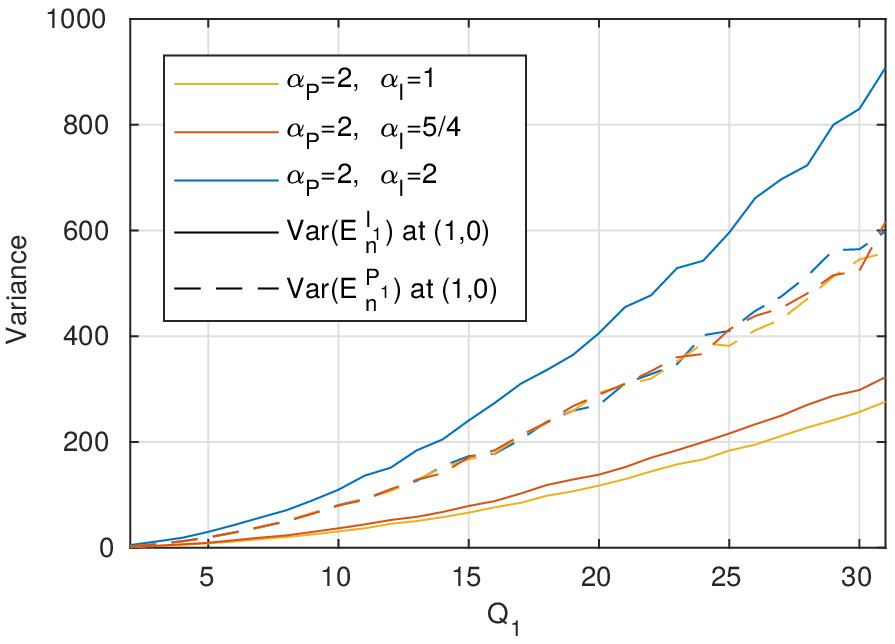}
\centerline{(b)}
\end{minipage}
\hfill
\begin{minipage}{0.32\linewidth}
\centering
\includegraphics[width=\linewidth]{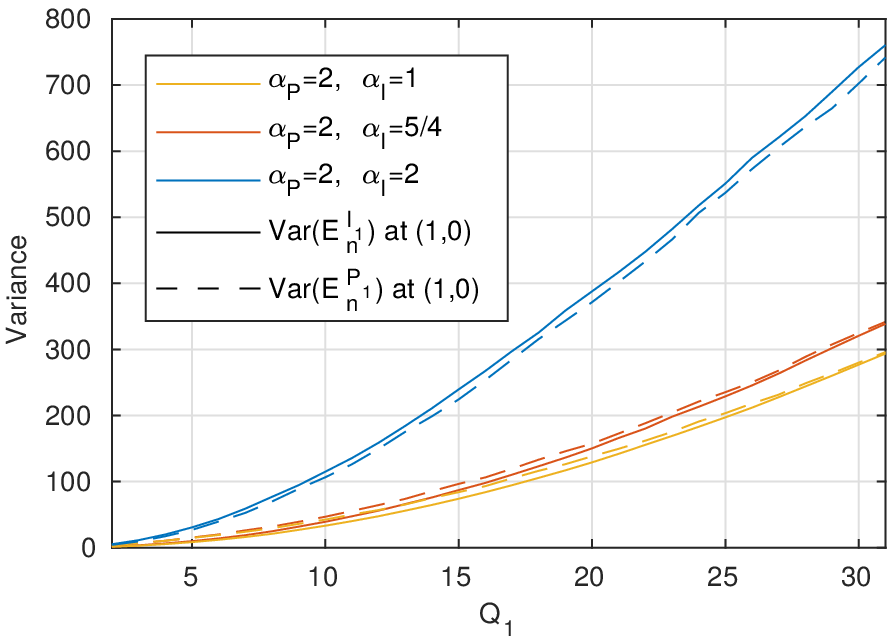}
\centerline{(c)}
\end{minipage}
\caption{Evolution of $\text{Var}(\mathbf{E}_n^{\text{I}_1})$ (\emph{solid}) and $\text{Var}(\mathbf{E}_n^{\text{P}_1})$ (\emph{dashed}) for $\alpha_{\text{P}}=2$ and varying $\alpha_{\text{I}}$ and $\text{Q}_1$: (a) static video (\texttt{akiyo}), (b) dynamic video (\texttt{mobile}), (c) synthetic model for static video ($\sigma_X^2=2500$).}
\label{fig:var_EI_EP}
\end{figure}

The above findings can be checked in Fig.~\ref{fig:var_EI_EP}, where the evolution of $\text{Var}\left(\mathbf{E}_n^{\text{I}_1}\right)$ and $\text{Var}\left(\mathbf{E}_n^{\text{P}_1}\right)$ for the $(1,0)$-th DCT coefficient is shown for two videos gathered from~\cite{XIPH}: the static video \emph{akiyo} in Fig.~\ref{fig:var_EI_EP}(a), and the dynamic video \emph{mobile} in Fig.~\ref{fig:var_EI_EP}(b). In the static case reported in Fig.~\ref{fig:var_EI_EP}(a), the evolution of $\text{Var}\left(\mathbf{E}_n^{\text{P}_1}\right)$ at $(1,0)$ mostly depends on $\alpha_{\text{I}}$ due to the role of the S-MBs that fundamentally make reference to MBs from the previously encoded I-frame, thus yielding a quantization error $\mathbf{E}_n^{\text{P}_1}$ very similar to $\mathbf{E}_n^{\text{I}_1}$. On the other hand, when dealing with dynamic scenes as in Fig.~\ref{fig:var_EI_EP}(b), $\text{Var}\left(\mathbf{E}_n^{\text{P}_1}\right)$ at $(1,0)$ does not depend on $\alpha_{\text{I}}$, instead it is governed by $\alpha_{\text{P}}$ given that the curves for different values of $\alpha_{\text{I}}$ are close to each other.

From the above analysis, it follows that for static video sequences, the difference $\text{Var}\left(\mathbf{E}_n^{\text{I}_1}\right)-\text{Var}\left(\mathbf{E}_n^{\text{P}_1}\right)$ is small and so $\text{Var}\left(\mathbf{W}_n\right)|_{n\in\mathsf{I}_1}\approx\text{Var}\left(\mathbf{W}_n\right)|_{n\in\mathsf{P}_1}$, while for dynamic videos it is harder to define a similar relation. In fact, the varying nature of the prediction residue with dynamic videos complicates the modeling (i.e., at least a motion estimation of the scene would be needed), thus we leave its study for a future work. In constrast, the nearly constant behavior of $\text{Var}\left(\mathbf{W}_n\right)$ under the intra prediction (independently of the type of frame used in the first compression), implies that for low-motion videos, the presence of the VPF is ultimately guided by the behavior of $\text{Var}\left(\mathbf{W}_n\right)$ under the inter prediction, which we analyze in the following Section~\ref{subsec:inter}.

Prior to address the evolution of the residue under the inter-prediction setting, we introduce the proposed semi-analytic model to predict the behavior of $\text{Var}\left(\mathbf{E}_n^{\text{I}_1}\right)$ and $\text{Var}\left(\mathbf{E}_n^{\text{P}_1}\right)$ for static video sequences. To that end, we use a first-order autoregressive process $x_n$ for modeling the temporal dependencies of the source signal $\mathbf{X}_n$. Troughout the paper we will consider three distinct time indices: $n-2$, $n-1$, and $n$, and we will correspondingly define three stochastic processes: $x_{n-2}$, $x_{n-1}$, and $x_{n}$. Thus, assuming the stochastic process $x_{n-2}$ follows a Laplacian distribution with zero mean and variance $\sigma_X^2=2500$, we generate the remaining stochastic processes as $x_{n-1}=\rho x_{n-2}+r_{n-1}$ and $x_n =\rho x_{n-1}+r_n$, where we take $\rho=0.99$ to simulate the high temporal correlation between adjacent frames in low-motion videos and we define the residue signal $r_n$ as a zero-mean Gaussian process with variance $\sigma_R^2=10$. On the other hand, the inter-prediction process is also characterized through a first-order autoregressive process, such that $\hat{\mathbf{X}}_n=\mathbf{X}_{n-1}^\prime(\mathbf{m})$ is modeled through the stochastic process $\hat{x}_n=\rho_{\text{P}}x_{n-1}^\prime+\nu_n$, where we take $\rho_{\text{P}}=0.88$ to simulate the effect of a non-perfect motion estimation and we make use of a zero-mean Gaussian process $\nu_n$ with variance $\sigma_{\nu}=1$ to mimic the effect of the motion compensation through $\mathbf{m}$. During the first compression, we assume that at time index $n-2$ there is always an I-frame, then at $n-1$ a P-frame, and finally at $n$, it will depend on the case under study, i.e., $n\in\mathsf{I}_1$ or $n\in\mathsf{P}_1$. For the second compression, an I-frame is also assumed at $n-2$, and a P-frame at $n-1$.

Now, from the definition of $\mathbf{E}_n^{\text{I}_1}$, we generate the samples of the corresponding synthetic signal as $e_n^{\text{I}_1}=x_n^\prime-x_n$, where all the samples from $x_n$ are first quantized as in \eqref{eq:u_q} for I-MBs (using $\Delta_1^{\text{I}}$ and $\alpha_{\text{I}}$) and then are accordingly reconstructed through \eqref{eq:u_prime} to obtain $x_n^\prime$. On the other hand, the synthetic signal for $\mathbf{E}_{n}^{\text{P}_1}$, must consider the different coding modes from $\mathcal{C}$ that can be used when a P-frame is encoded. Without loss of generality, we assume that the number of I-MBs in static videos is negligible, such that $\text{p}(\text{I-MB})\rightarrow0$, which is commonly the case in practice. Regarding the modeling of P-MBs, we assume that $\text{p}(\text{P-MB})$ decreases as $\text{Q}_1$ grows (which is the expected behavior in low-motion videos), so we use $\text{p}(\text{P-MB})=p_1$ with $p_1\triangleq0.15 + 0.7e^{-9\frac{\text{Q}_1}{\text{Q}_{\max}}}$, where $\text{Q}_{\max}=31$ denotes the maximum allowed quantization parameter. Accordingly, we set $\text{p}(\text{S-MB})=1-p_1$ and we finally compute the synthetic signal for $\mathbf{E}_{n}^{\text{P}_1}$ as
\begin{equation*}
e_{n}^{\text{P}_1}=\begin{cases}
u_{n}^\prime-u_{n}, & \text{if }\text{c}=\text{P-MB}\\
x_{n-1}^\prime-x_{n}, & \text{otherwise}\\
\end{cases},
\end{equation*}
where $u_n=x_n-\hat{x}_n$ and $u_{n}^\prime$ is obtained by first quantizing $u_n$ as in the second case of \eqref{eq:u_q} and then reconstructing its samples through \eqref{eq:u_prime} (in both cases, using $\Delta_1^{\text{P}}$ and $\alpha_{\text{P}}$). Finally, to obtain $x_{n-1}^\prime$, the whole encoding process of a P-frame must be applied, so with probability $1-p_1$, the samples of $x_{n-1}^\prime$ stem from the use of the first case in \eqref{eq:u_q}-\eqref{eq:u_prime} over the samples of $x_{n-1}$ (using $\Delta_1^{\text{I}}$ and $\alpha_{\text{I}}$). On the other hand, with probability $p_1$, the corresponding signal $u_{n-1}=x_{n-1}-\hat{x}_{n-1}$ must be computed, where $\hat{x}_{n-1}$ is given through $\hat{x}_{n-1}=\rho_{\text{P}}x_{n-2}^\prime+\nu_{n-1}$ with $x_{n-2}^\prime$ being the result of reconstructing the whole signal $x_{n-2}$ as the encoding of an I-frame. Once $u_{n-1}$ is computed, then the second case in \eqref{eq:u_q} and \eqref{eq:u_prime} must be applied to generate $u_{n-1}^\prime$, which finally allows the calculation of $x_{n-1}^\prime=u_{n-1}^\prime+\hat{x}_{n-1}$.

The result of using this semi-analytic approach to predict the behavior of $\text{Var}\left(\mathbf{E}_n^{\text{I}_1}\right)$ and $\text{Var}\left(\mathbf{E}_n^{\text{P}_1}\right)$ in the DCT domain at $(1,0)$ for static video sequences is illustrated in Fig.~\ref{fig:var_EI_EP}(c), which closely resembles its empirical counterpart in Fig.~\ref{fig:var_EI_EP}(a).

\subsection{Inter-prediction residue analysis}
\label{subsec:inter}

In this case, $\hat{\mathbf{Y}}_n$ is the result of an inter prediction, i.e., $\hat{\mathbf{Y}}_n=\mathbf{Y}_{n-1}^\prime(\mathbf{m})$. Assuming that the estimated motion scene through $\mathbf{m}$ coincides in the two consecutive compressions (which is reasonable in practice, provided that the content of the scene remains unchanged across both compressions), $\hat{\mathbf{Y}}_n$ can be expressed as
\begin{align}
\hat{\mathbf{Y}}_n &=\mathbf{Y}_{n-1}(\mathbf{m})+\mathbf{E}_{n-1}^{\text{P}_2}\nonumber\\
&=\mathbf{X}_{n-1}(\mathbf{m})+\mathbf{E}_{n-1}^{\text{P}_1}+\mathbf{E}_{n-1}^{\text{P}_2},
\label{eq:hat_Y_n}
\end{align}
where $\mathbf{E}_{n-1}^{\text{P}_1}\triangleq\mathbf{U}_{n-1}^\prime(\mathbf{m})-\mathbf{U}_{n-1}(\mathbf{m})$ and $\mathbf{E}_{n-1}^{\text{P}_2}\triangleq\mathbf{W}_{n-1}^\prime(\mathbf{m})-\mathbf{W}_{n-1}(\mathbf{m})$ represent the quantization errors that result from the first and second compression, respectively. Using \eqref{eq:Y_n} and \eqref{eq:hat_Y_n} in the definition of $\mathbf{W}_n$, we obtain 
\begin{equation*}
\mathbf{W}_n=\mathbf{R}_n+\mathbf{E}_n, \quad\text{with}\quad
\mathbf{E}_n=\begin{cases}
\mathbf{E}_n^{\text{I}_1}-\mathbf{E}_{n-1}^{\text{P}_1}-\mathbf{E}_{n-1}^{\text{P}_2}, & \text{if } n\in\mathsf{I}_1\\
\mathbf{E}_n^{\text{P}_1}-\mathbf{E}_{n-1}^{\text{P}_1}-\mathbf{E}_{n-1}^{\text{P}_2}, & \text{if } n\in\mathsf{P}_1\end{cases},
\end{equation*}
where $\mathbf{R}_n\triangleq \mathbf{X}_n-\mathbf{X}_{n-1}(\mathbf{m})$ represents the prediction residue without any quantization error and $\mathbf{E}_n$ comprises all the quantization errors that emerge during the two successive compressions. Now, assuming that these quantization errors have negligible correlation with $\mathbf{R}_n$, we can approximate the variance of $\mathbf{W}_n$ as 
\begin{equation*}
\text{Var}\left(\mathbf{W}_n\right) = \text{Var}\left(\mathbf{R}_n\right) + \text{Var}\left(\mathbf{E}_n\right).
\end{equation*}
Therefore, in this case, \eqref{eq:W_n} becomes $\text{Var}\left(\mathbf{E}_n\right)|_{n\in\mathsf{I}_1}-\text{Var}\left(\mathbf{E}_n\right)|_{n\in\mathsf{P}_1}$, which after deriving the expression for $\text{Var}\left(\mathbf{E}_n\right)$, can be expressed as
\begin{align}
\text{Var}\left(\mathbf{E}_n\right)&|_{n\in\mathsf{I}_1}-\text{Var}\left(\mathbf{E}_n\right)|_{n\in\mathsf{P}_1}\nonumber\\
&=\underbrace{\text{Var}\left(\mathbf{E}_n^{\text{I}_1}\right)-\text{Var}\left(\mathbf{E}_n^{\text{P}_1}\right)-2\left(\text{cov}\left(\mathbf{E}_n^{\text{I}_1},\mathbf{E}_{n-1}^{\text{P}_1}\right)-\text{cov}\left(\mathbf{E}_n^{\text{P}_1},\mathbf{E}_{n-1}^{\text{P}_1}\right)\right)}_{\text{depends on $\text{Q}_1$}}\nonumber\\
&\quad\underbrace{-2\left(\text{cov}\left(\mathbf{E}_n^{\text{I}_1},\mathbf{E}_{n-1}^{\text{P}_2}\right)-\text{cov}\left(\mathbf{E}_n^{\text{P}_1},\mathbf{E}_{n-1}^{\text{P}_2}\right)\right)}_{\text{depends on $\text{Q}_1$ and $\text{Q}_2$}}.
\label{eq:diff_sigma_E}
\end{align}
In the above equation, some of the terms uniquely depend on the quantization parameter used during the first compression $\text{Q}_1$, whereas the rest depends on the two quantization parameters $\text{Q}_1$ and $\text{Q}_2$ applied in the double compression scheme. We first center our attention on the terms that only depend on $\text{Q}_1$, except for $\text{Var}\left(\mathbf{E}_n^{\text{I}_1}\right)$ and $\text{Var}\left(\mathbf{E}_n^{\text{P}_1}\right)$, which have already been described in Section~\ref{subsec:intra}. Without loss of generality, we analyze the behavior of the two covariance functions through their corresponding correlations:\footnote{Note that for two arbitrary random variables $A$ and $B$, their covariance and correlation relate as follows: $\text{corr}\left(A,B\right)=\text{cov}\left(A,B\right)/(\text{Var}(A)\text{Var}(B))$.}

\begin{enumerate}
\item $\text{cov}\left(\mathbf{E}_n^{\text{I}_1},\mathbf{E}_{n-1}^{\text{P}_1}\right)$: as hinted in Section~\ref{subsec:intra}, because of the large amount of S-MBs that show up in static videos, the correlation between $\mathbf{E}_n^{\text{I}_1}$ and $\mathbf{E}_{n-1}^{\text{P}_1}$ is expected to be significant. Moreover, the larger the value of $\text{Q}_1$, the higher becomes the number of S-MBs and, as a consequence, $\text{corr}\left(\mathbf{E}_n^{\text{I}_1},\mathbf{E}_{n-1}^{\text{P}_1}\right)$ grows with $\text{Q}_1$. This behavior can be observed in Fig.~\ref{fig:corr_EI_E2_EP_E2}(a), where the empirical correlation measured from the low-motion video \texttt{akiyo} is depicted. The largest correlation value is achieved when $\alpha_{\text{I}}=\alpha_{\text{P}}$ and its value proportionally decreases as $\alpha_{\text{I}}$ moves away from $\alpha_{\text{P}}$.

In this case, the semi-analytic model for $\mathbf{E}_n^{\text{I}_1}$ and $\mathbf{E}_{n-1}^{\text{P}_1}$ is derived in the same way as discussed in Section~\ref{subsec:intra} with the peculiarity that now the latter quantization error is obtained at $n-1$ instead of $n$, but the same procedure is followed and the same values of $p_1$ are employed. By doing so, the correlation between the derived synthetic signals $e_n^{\text{I}_1}$ and $e_{n-1}^{\text{P}_1}$ is depicted in Fig.~\ref{fig:corr_EI_E2_EP_E2}(b), where it can be observed that the proposed synthetic model does not follow very well the empirical cases when $\alpha_{\text{I}}\in\{1,\frac{5}{4}\}$, so probably an adjustment of the model parameters would be needed in these cases.

\begin{figure}
\centering
\begin{minipage}{0.49\linewidth}
\centering
\includegraphics[width=\linewidth]{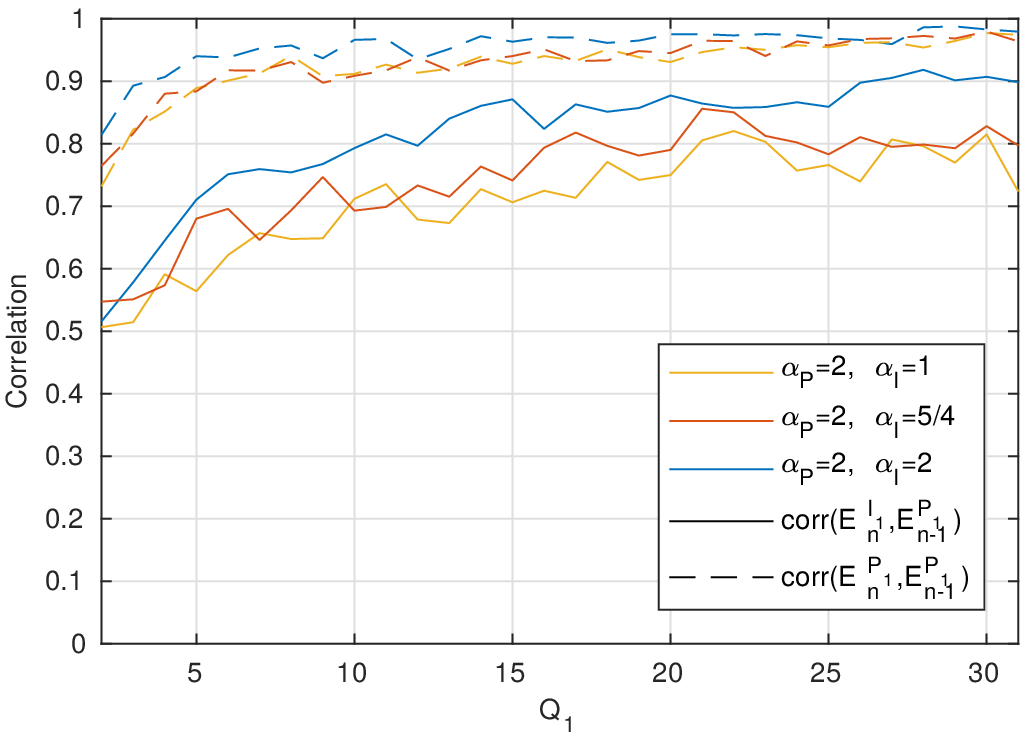}
\centerline{(a)}
\end{minipage}
\hfill
\begin{minipage}{0.49\linewidth}
\centering
\includegraphics[width=\linewidth]{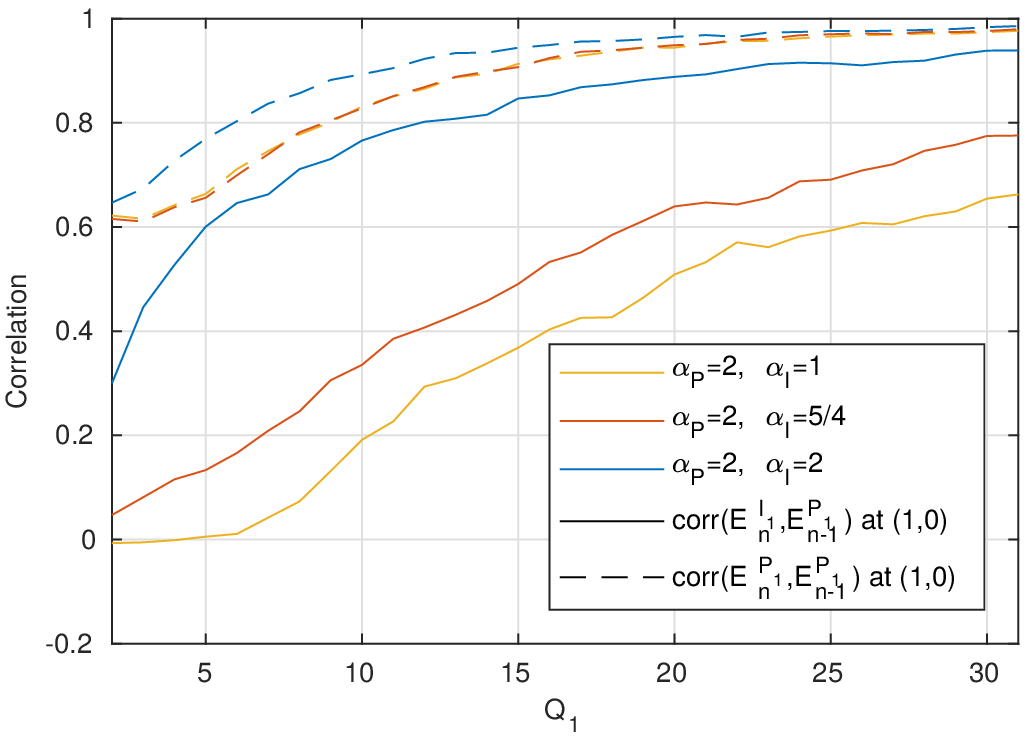}
\centerline{(b)}
\end{minipage}
\caption{Evolution of $\text{corr}\left(\mathbf{E}_n^{\text{I}_1},\mathbf{E}_{n-1}^{\text{P}_1}\right)$ (\emph{solid lines}) and $\text{corr}\left(\mathbf{E}_n^{\text{P}_1},\mathbf{E}_{n-1}^{\text{P}_1}\right)$ (\emph{dashed lines}) as a function of $\alpha_{\text{I}}$, $\alpha_{\text{P}}$, and $\text{Q}_1$: (a) \texttt{akiyo}, (b) synthetic model.}
\label{fig:corr_EI_E2_EP_E2}
\end{figure}

\item $\text{cov}\left(\mathbf{E}_n^{\text{P}_1},\mathbf{E}_{n-1}^{\text{P}_1}\right)$: the corresponding correlation is computed between the quantization errors that arise from two inter-predictive frames at different time indices, i.e., $\mathbf{E}_n^{\text{P}_1}$ and $\mathbf{E}_{n-1}^{\text{P}_1}$, during the first compression. Given that $\mathbf{U}_n$ is very close to $\mathbf{U}_{n-1}$ in static scenes, we expect a very high correlation in this case, which is empirically confirmed in Fig.~\ref{fig:corr_EI_E2_EP_E2}(a). 

Fig.~\ref{fig:corr_EI_E2_EP_E2}(b) collects the resulting correlation after generating the synthetic versions of the quantization errors $\mathbf{E}_n^{\text{P}_1}$ and $\mathbf{E}_{n-1}^{\text{P}_1}$. Differently from the previous point, now the synthetically computed correlations follow very well the trend of the empirical ones, except for small values of $\text{Q}_1$.
\end{enumerate}

The analysis in Section~\ref{subsec:intra} of the variance terms in \eqref{eq:diff_sigma_E} reveals that for low-motion videos the value of $\text{Var}\left(\mathbf{E}_n^{\text{I}_1}\right)-\text{Var}\left(\mathbf{E}_n^{\text{P}_1}\right)$ is generally small. Hence, its effect is negligible in the evolution of \eqref{eq:diff_sigma_E}. Additionally, since $\text{Var}\left(\mathbf{E}_n^{\text{I}_1}\right)\approx\text{Var}\left(\mathbf{E}_n^{\text{P}_1}\right)$, this implies that $\text{cov}\left(\mathbf{E}_n^{\text{I}_1},\mathbf{E}_{n-1}^{\text{P}_1}\right)-\text{cov}\left(\mathbf{E}_n^{\text{P}_1},\mathbf{E}_{n-1}^{\text{P}_1}\right)$ is proportional to the difference of the corresponding correlations. From the example shown in Fig.~\ref{fig:corr_EI_E2_EP_E2}(a), one can observe that such difference is nearly constant for distinct values of $\text{Q}_1$ independently of the considered relation between $\alpha_{\text{I}}$ and $\alpha_{\text{P}}$, so we can ensure that the terms dependent on $\text{Q}_1$ do not cause prominent changes in \eqref{eq:diff_sigma_E} for low-motion videos. As a consequence, the appearance of the VPF is fundamentally determined by the two covariance terms that jointly depend on the quantization parameters $\text{Q}_1$ and $\text{Q}_2$, which we describe below:

\begin{enumerate}
\item $\text{cov}\left(\mathbf{E}_n^{\text{I}_1},\mathbf{E}_{n-1}^{\text{P}_2}\right)$: in this case, we need to consider the correlation between the quantization errors that arise during two distinct compression stages: $\mathbf{E}_n^{\text{I}_1}$ during the first compression, whose synthetic modeling has already been detailed in Section~\ref{subsec:intra}), and $\mathbf{E}_{n-1}^{\text{P}_2}$ during the second stage. The synthetic signal for $\mathbf{E}_{n-1}^{\text{P}_2}$ can be derived following a similar procedure to that of $\mathbf{E}_{n}^{\text{P}_1}$ (described in Section~\ref{subsec:intra}), but taking now into account that the input signal has been compressed once and that the recompression of a video sequence alters the probabilities of each MB type. In fact, the probability of having a P-MB during the second compression is now lowered to $p_2 \triangleq 0.15 + (p1-0.15)e^{-9\frac{\text{Q}_2}{\text{Q}_{\max}}}$ because in double compressed videos we expect to find prediction residues closer to zero as an effect of the first compression, and so we count on finding a smaller number of P-MBs. Hence, using $\text{p}(\text{P-MB})=p_2$ and $\text{p}(\text{S-MB})=1-p_2$, we compute the synthetic signal as 
\begin{equation*}
e_{n-1}^{\text{P}_2}=\begin{cases}
w_{n-1}^\prime-w_{n-1}, & \text{if }\text{c}=\text{P-MB}\\
y_{n-2}^\prime-y_{n-1}, & \text{otherwise}\\
\end{cases},
\end{equation*}
where $w_{n-1}=x_{n-1}^\prime-\hat{y}_{n-1}$, $\hat{y}_{n-1}=\rho_{\text{P}}x_{n-2}^\prime+\nu_{n-1}$, and $w_{n-1}^\prime$ is obtained by first quantizing $w_{n-1}$ as in the second case of \eqref{eq:u_q} and then reconstructing its samples through \eqref{eq:u_prime} (in both cases, using $\Delta_2^{\text{P}}$ and $\alpha_{\text{P}}$). On the other hand, we have that $y_{n-1}=x^\prime_{n-1}$ (see Sect.~\ref{subsec:intra}) and also that $y_{n-2}^\prime$ is the result of reconstructing the whole signal $x_{n-2}^\prime$ as the encoding of an I-frame with $\text{Q}_2$.

To check the validity of the model, we compare in Fig.~\ref{fig:corr_EI_E3} the obtained synthetic results of $\text{corr}\left(\mathbf{E}_n^{\text{I}_1},\mathbf{E}_{n-1}^{\text{P}_2}\right)$ fixing $\alpha_{\text{P}}$ and varying the values of $\alpha_{\text{I}}$, $\text{Q}_1$, and $\text{Q}_2$ (\emph{upper panels}) with the ones stemming from video sequence \texttt{akiyo} (\emph{center panels}). The similarity among the depicted results supports the validity of the model. Although it is hard to predict the exact correlation between two quantization errors of different nature, we give some hints on why there are regions in the correlation maps shown in Fig.~\ref{fig:corr_EI_E3} that share the same correlation sign. Let us consider a sample located at the same position in $\mathbf{E}_n^{\text{I}_1}$ and $\mathbf{E}_{n-1}^{\text{P}_2}$. As long as the sample sign is retained in the succeeding compressions, this sample contributes positively in the resulting correlation. However, a change of sign across both compressions yields a negative contribution. Therefore, the regions in the correlation map with the same sign indicate that under these cases the two quantization errors share the same direction.

\begin{figure}[t]
\centering
\begin{minipage}{0.32\linewidth}
\centering
\includegraphics[width=\linewidth]{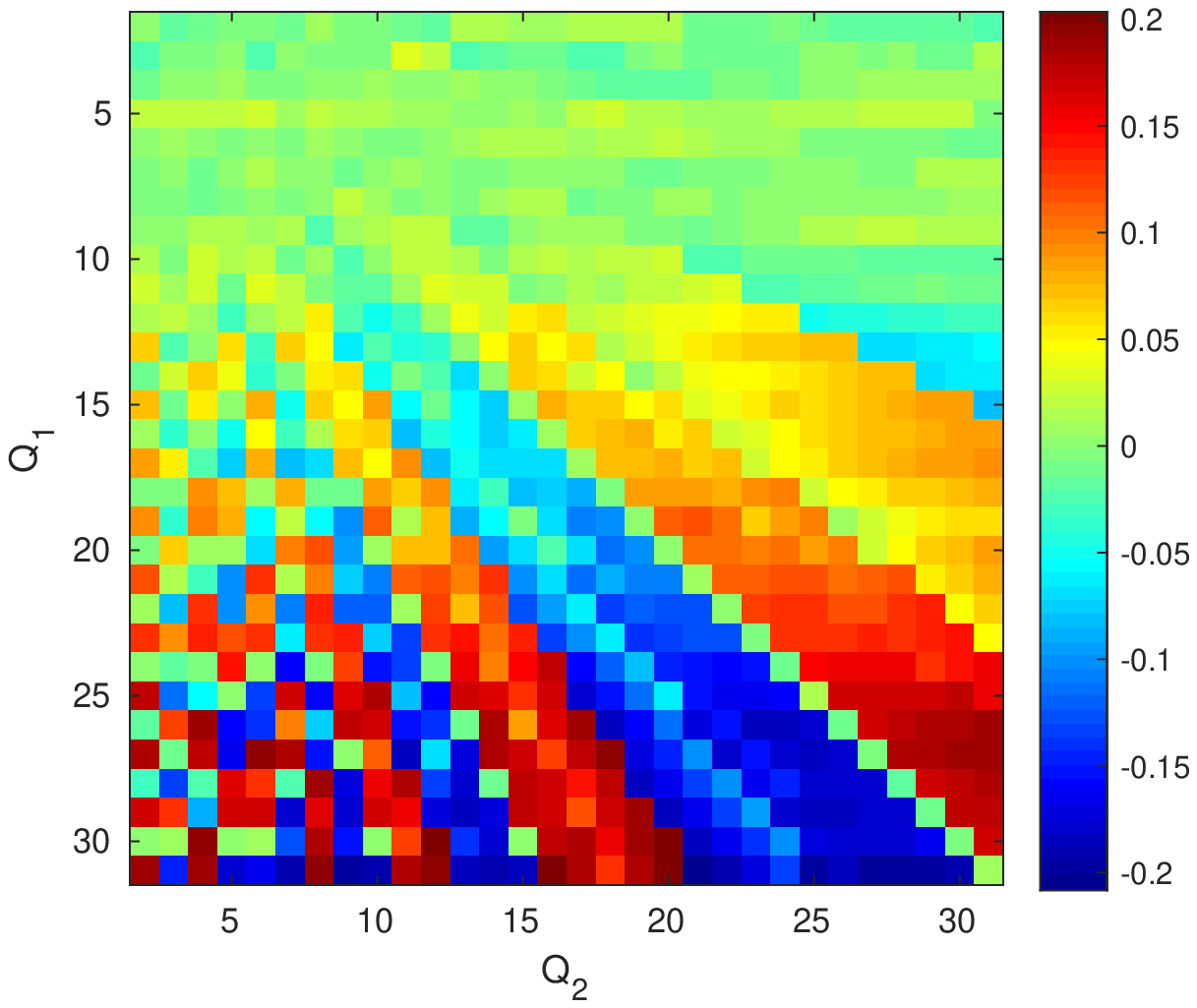}
\centerline{(a1)}
\end{minipage}
\begin{minipage}{0.32\linewidth}
\centering
\includegraphics[width=\linewidth]{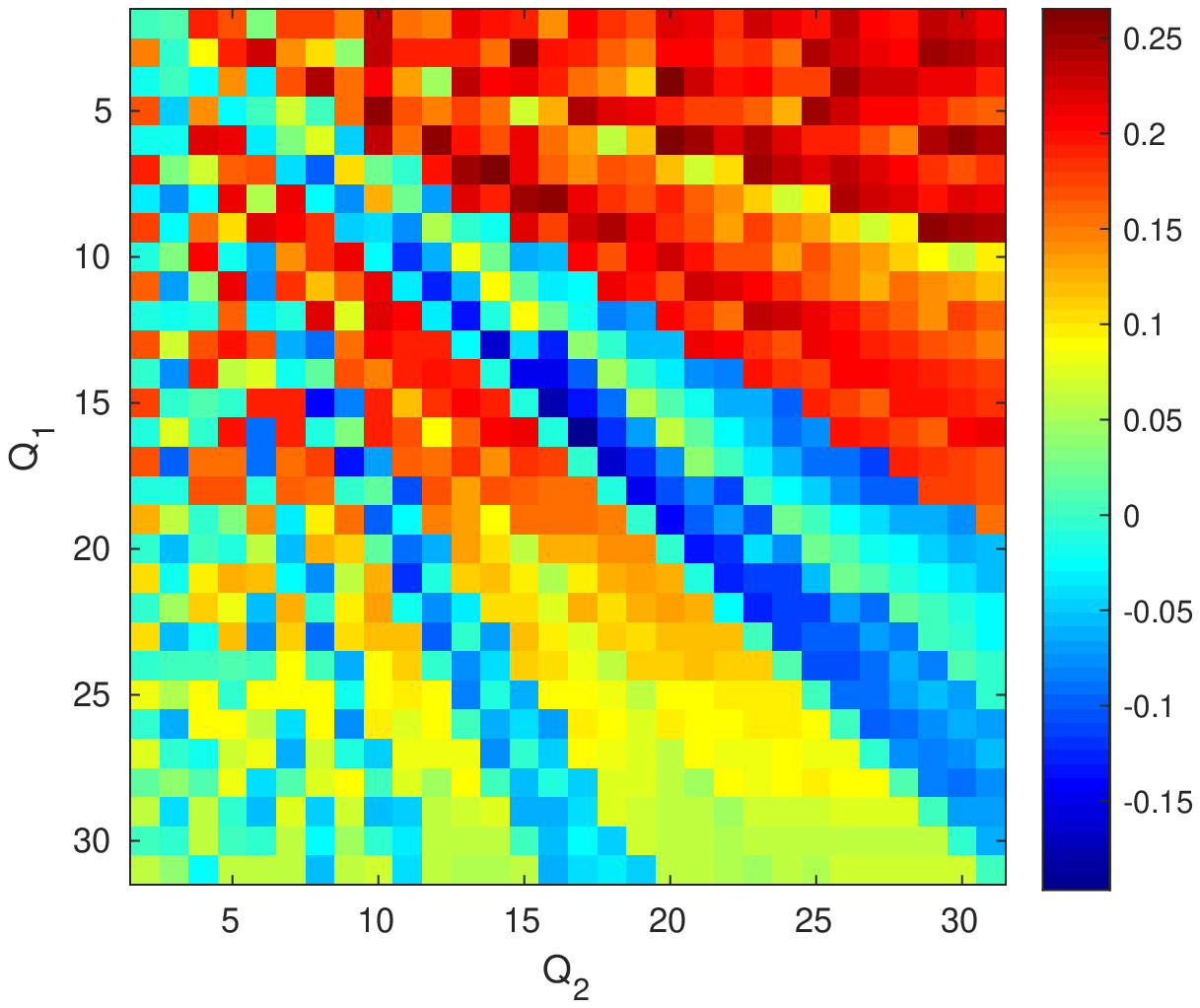}
\centerline{(b1)}
\end{minipage}
\begin{minipage}{0.32\linewidth}
\centering
\includegraphics[width=\linewidth]{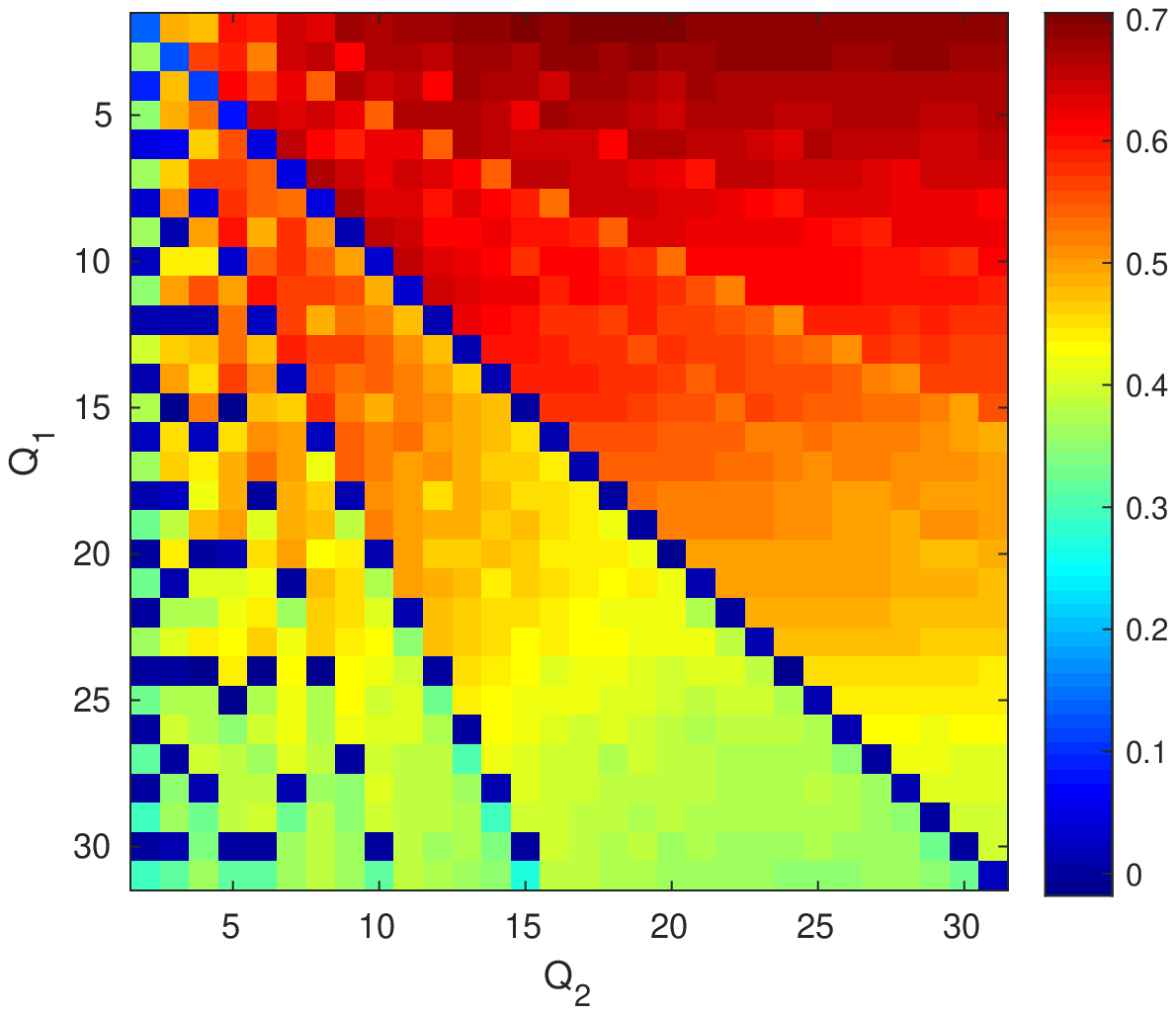}
\centerline{(c1)}
\end{minipage}
\\
\begin{minipage}{0.32\linewidth}
\centering
\includegraphics[width=\linewidth]{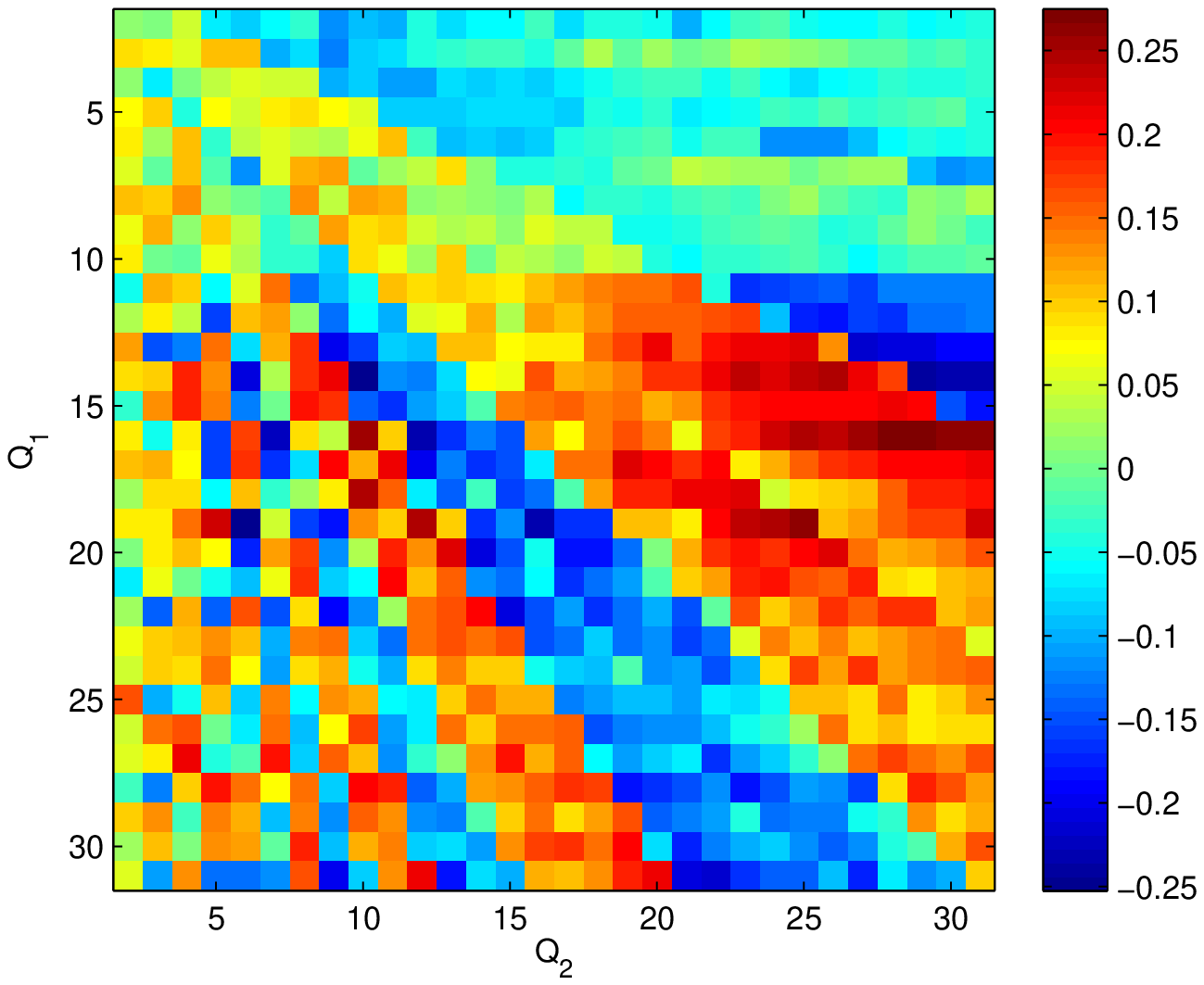}
\centerline{(a2)}
\end{minipage}
\begin{minipage}{0.32\linewidth}
\centering
\includegraphics[width=\linewidth]{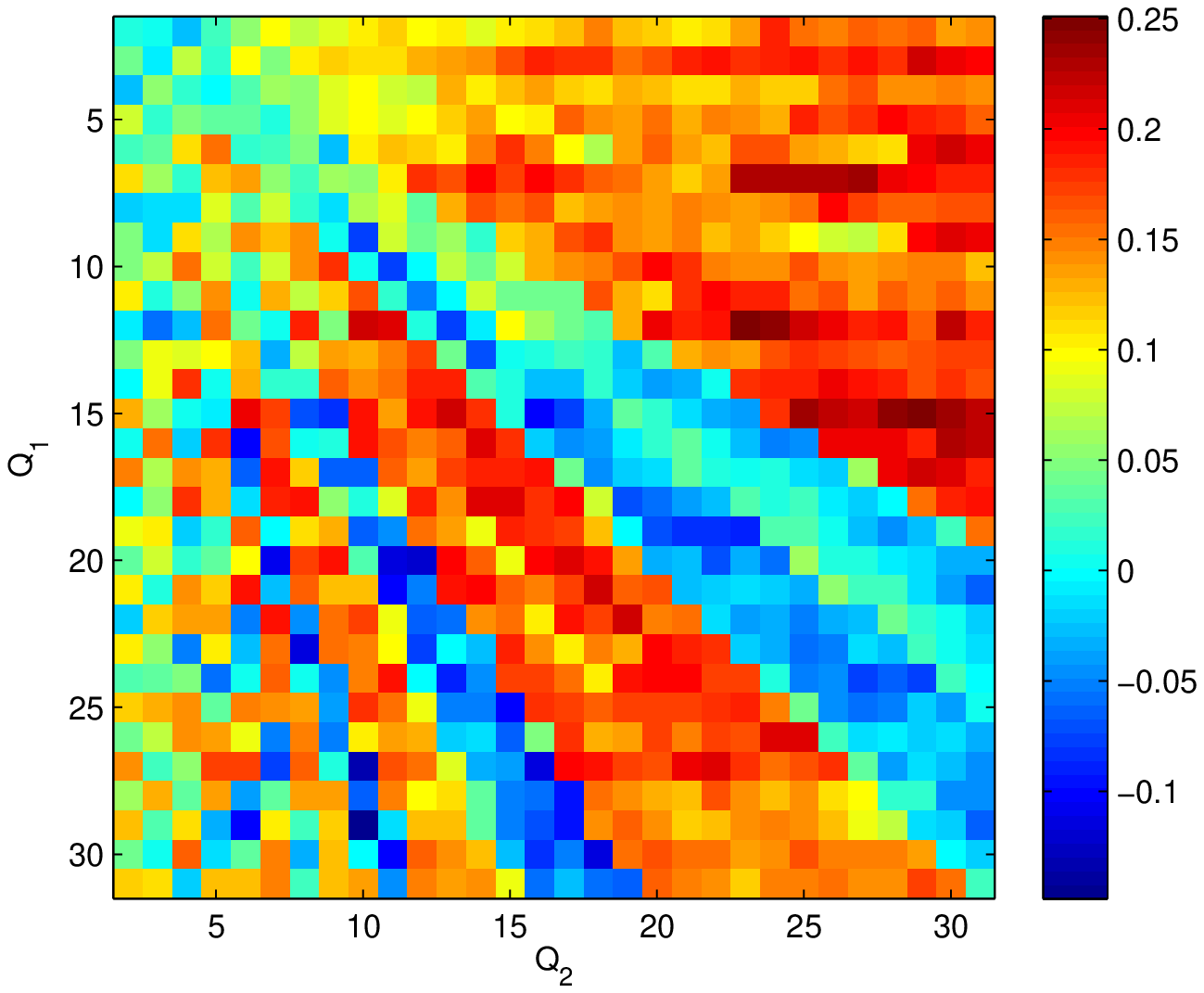}
\centerline{(b2)}
\end{minipage}
\begin{minipage}{0.32\linewidth}
\centering
\includegraphics[width=\linewidth]{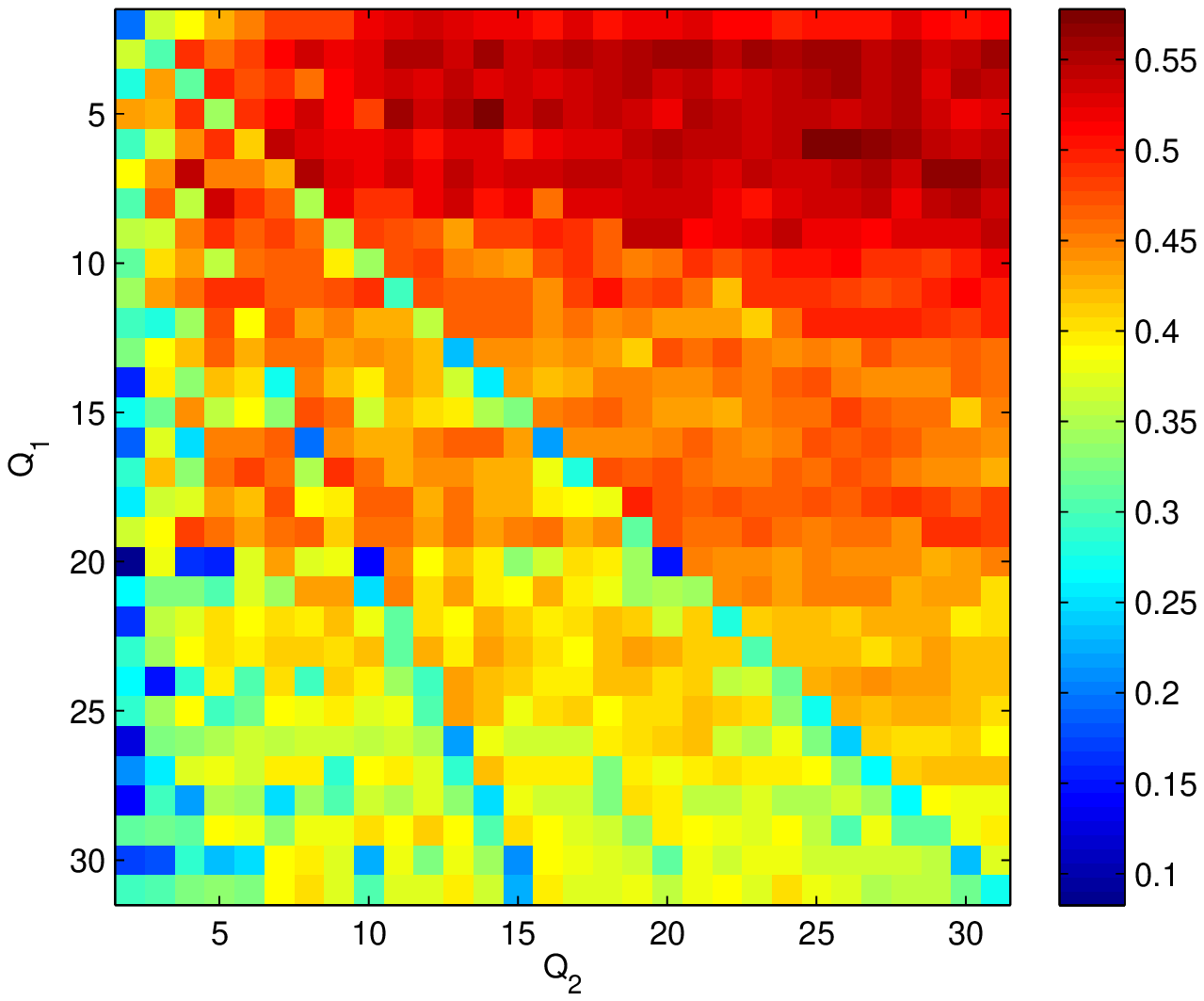}
\centerline{(c2)}
\end{minipage}
\\
\begin{minipage}{0.32\linewidth}
\centering
\includegraphics[width=\linewidth]{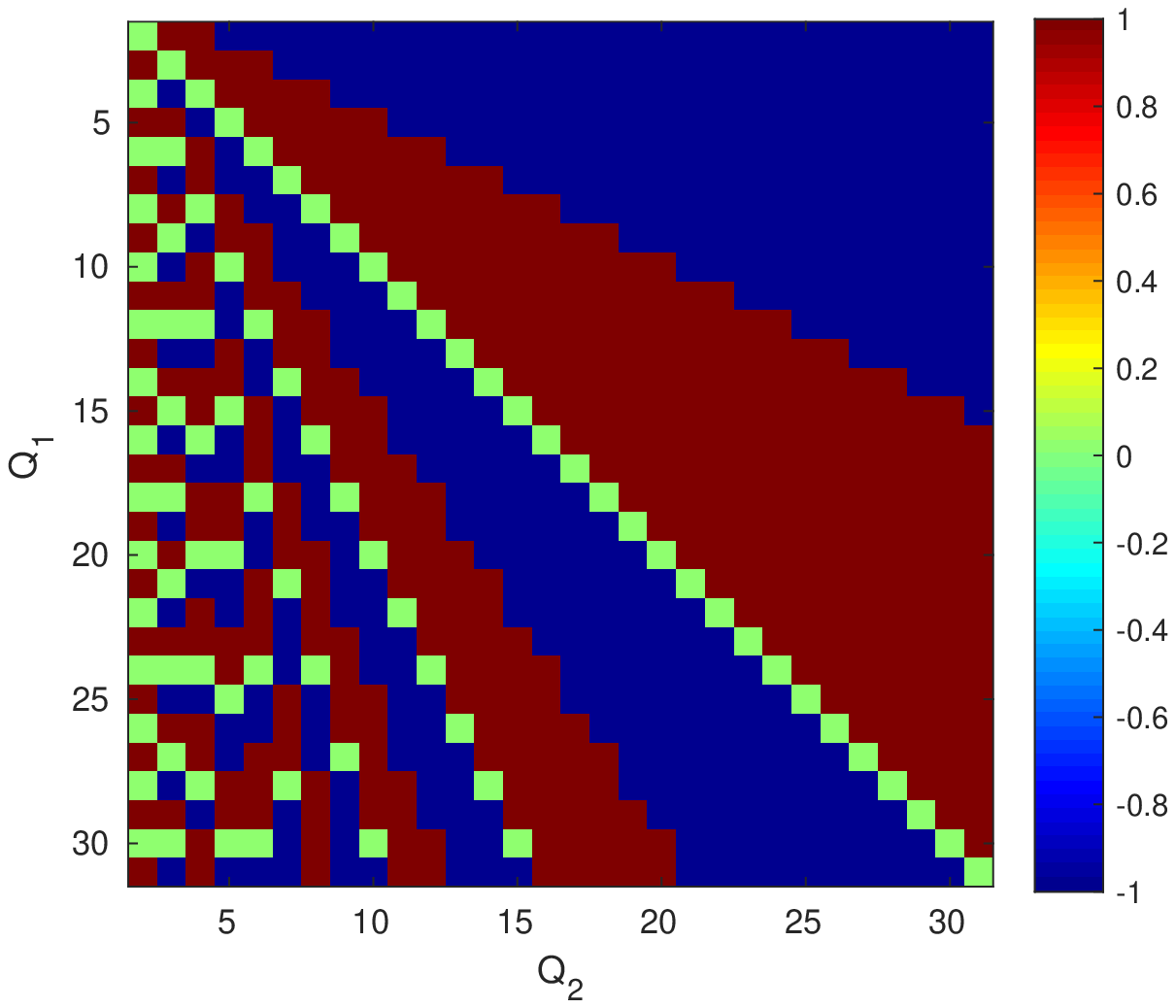}
\centerline{(a3)}
\end{minipage}
\begin{minipage}{0.32\linewidth}
\centering
\includegraphics[width=\linewidth]{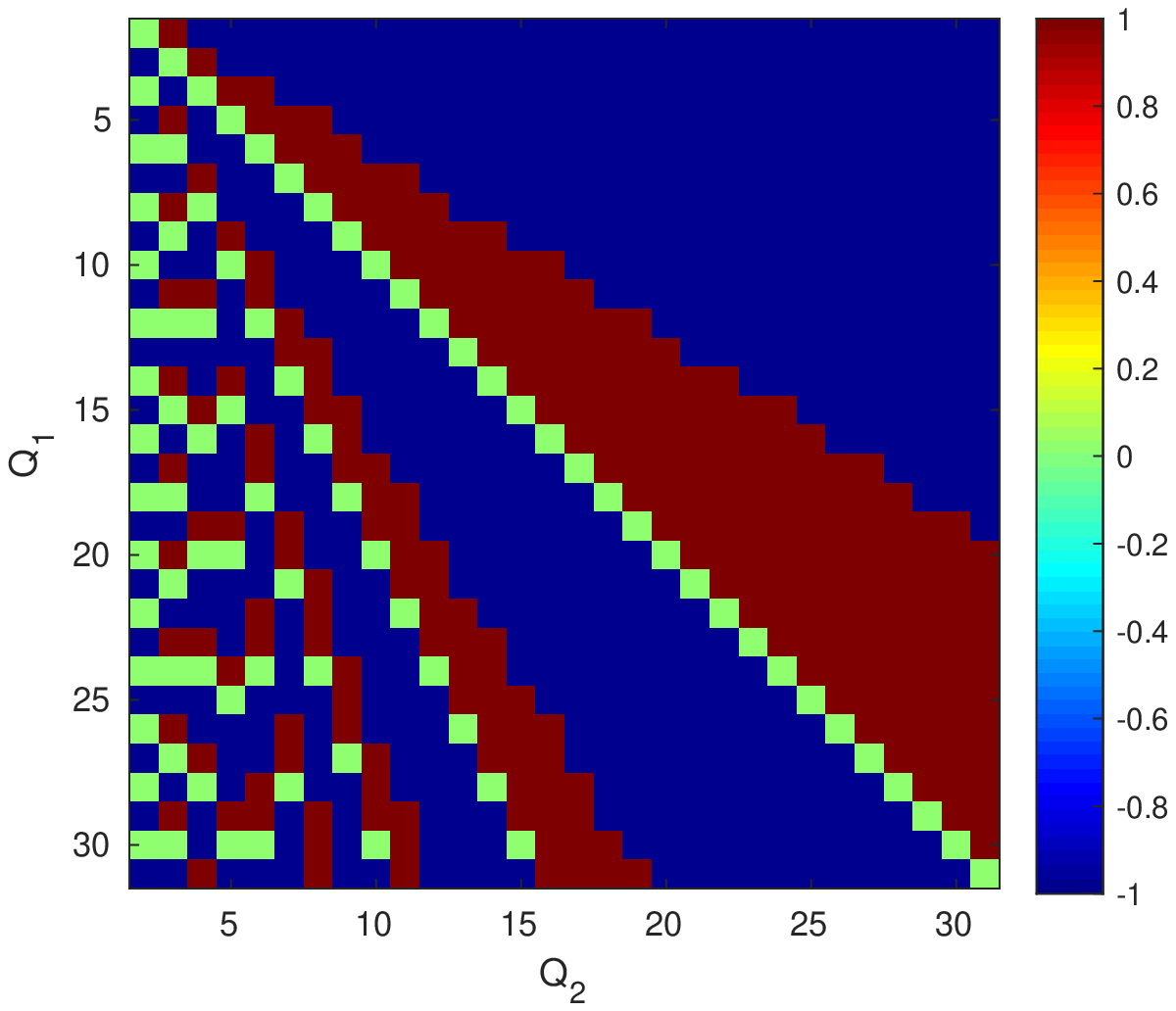}
\centerline{(b3)}
\end{minipage}
\begin{minipage}{0.32\linewidth}
\centering
\includegraphics[width=\linewidth]{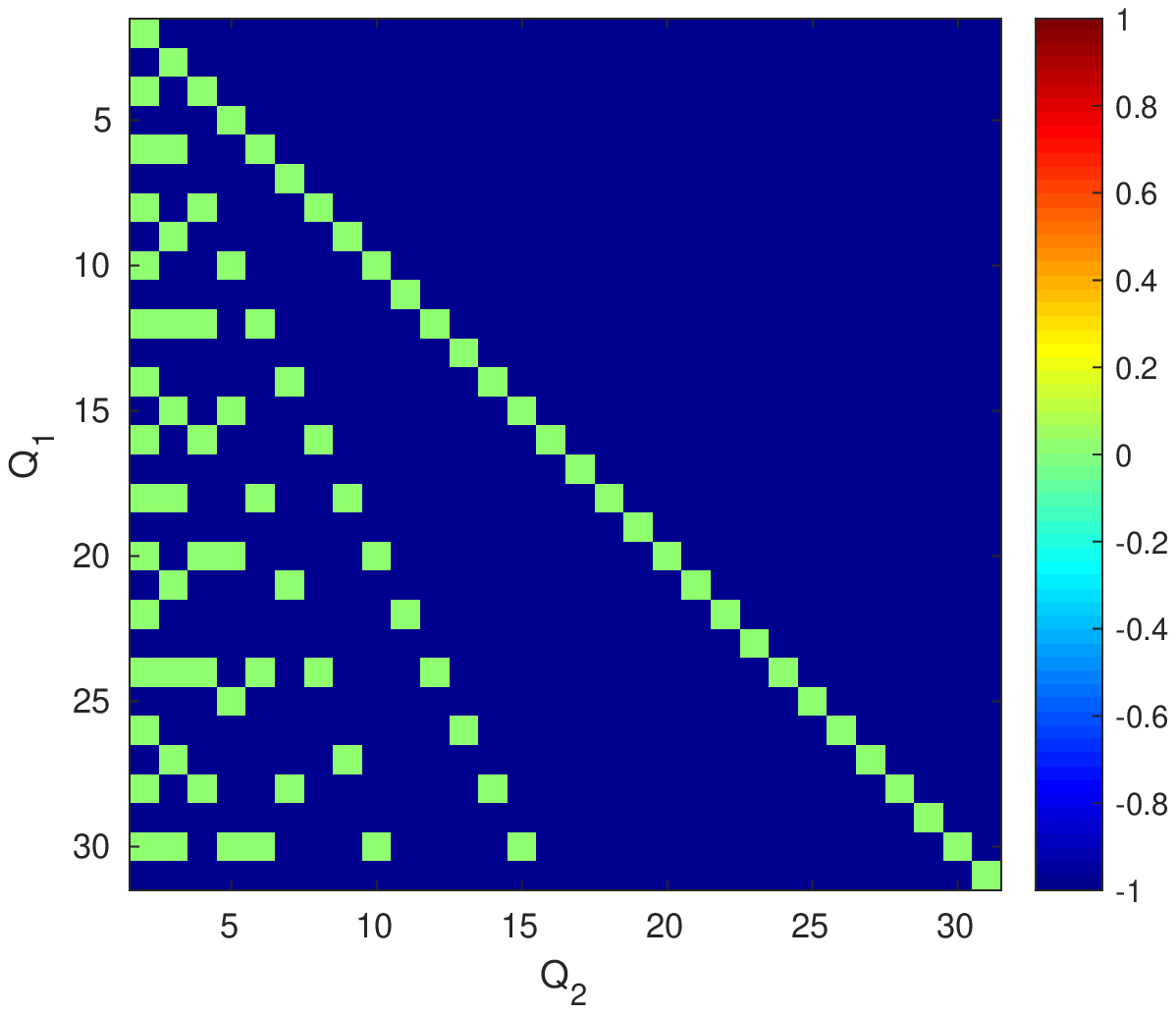}
\centerline{(c3)}
\end{minipage}
\caption{Evolution of $\text{corr}\left(\mathbf{E}_n^{\text{I}_1},\mathbf{E}_{n-1}^{\text{P}_2}\right)$ for a fixed $\alpha_{\text{P}}=2$ and varying $\alpha_{\text{I}}$, $\text{Q}_1$, and $\text{Q}_2$. The upper panels show the obtained results with synthetic signals, while the center panels correspond to the static video \texttt{akiyo}. In the lower panels, only the sign of the correlation is shown for those samples that have been quantized to the centroid $\Delta_1^{\text{I}}$ in the first compression stage. (a1-a3) $\alpha_{\text{I}}=1$, (b1-b3) $\alpha_{\text{I}}=\frac{5}{4}$, (c1-c3) $\alpha_{\text{I}}=2$.}
\label{fig:corr_EI_E3}
\end{figure}

Now, what delimits the particular shape of such regions is the relation between the selected width for the quantizer deadzones and the different reconstruction procedures that result in $\mathbf{E}_n^{\text{I}_1}$ and $\mathbf{E}_{n-1}^{\text{P}_2}$. As an example, we show in the lower panels of Fig.~\ref{fig:corr_EI_E3} the evolution of $\text{sgn}\left(d^\prime-d\right)$, where we set $d=\Delta_1^{\text{I}}$ and we obtain $d^\prime$ as its reconstructed version after quantizing $d$ with $\Delta_2^{\text{I}}$ (the rule for I-MBs is followed in both cases). This basic example shows some of the particular shapes that arise in the mentioned correlation maps and reflects, for instance, that in Fig.~\ref{fig:corr_EI_E3}(c3) no change of sign occurs for $\text{Q}_2>\text{Q}_1$ and $\alpha_{\text{I}}=\alpha_{\text{P}}=2$, while in Figs.~\ref{fig:corr_EI_E3}(a3) and~\ref{fig:corr_EI_E3}(b3) the flip of sign takes place at $\text{Q}_2>(2/\alpha_{\text{I}})\text{Q}_1$. This last relation will determine the limit beyond which the strength of the VPF vanishes for $\alpha_{\text{I}}\in\{1,\frac{5}{4}\}$.

\begin{figure*}[ht!]
\centering
\begin{minipage}{0.32\linewidth}
\centering
\includegraphics[width=\linewidth]{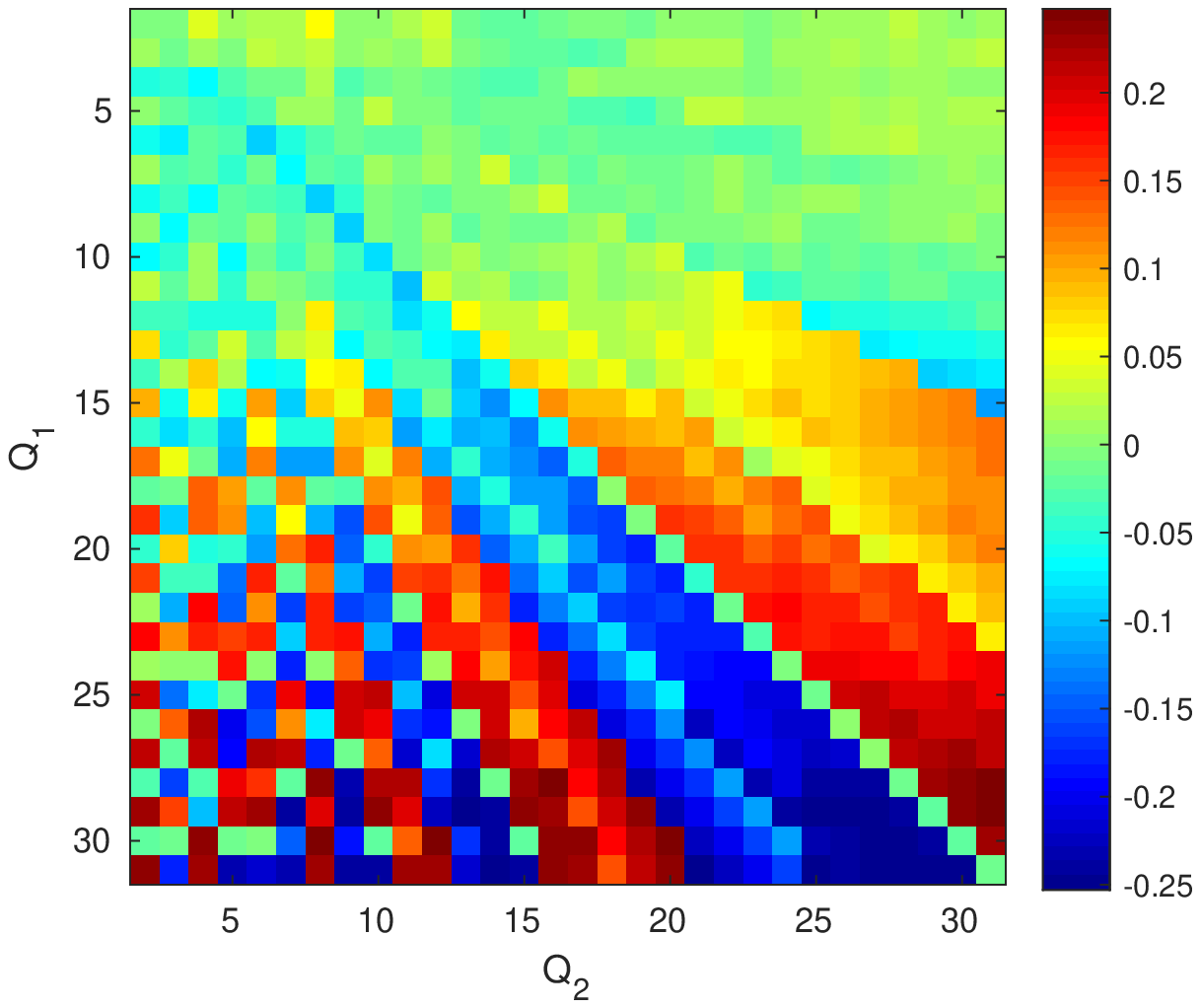}
\centerline{(a1)}
\end{minipage}
\begin{minipage}{0.32\linewidth}
\centering
\includegraphics[width=\linewidth]{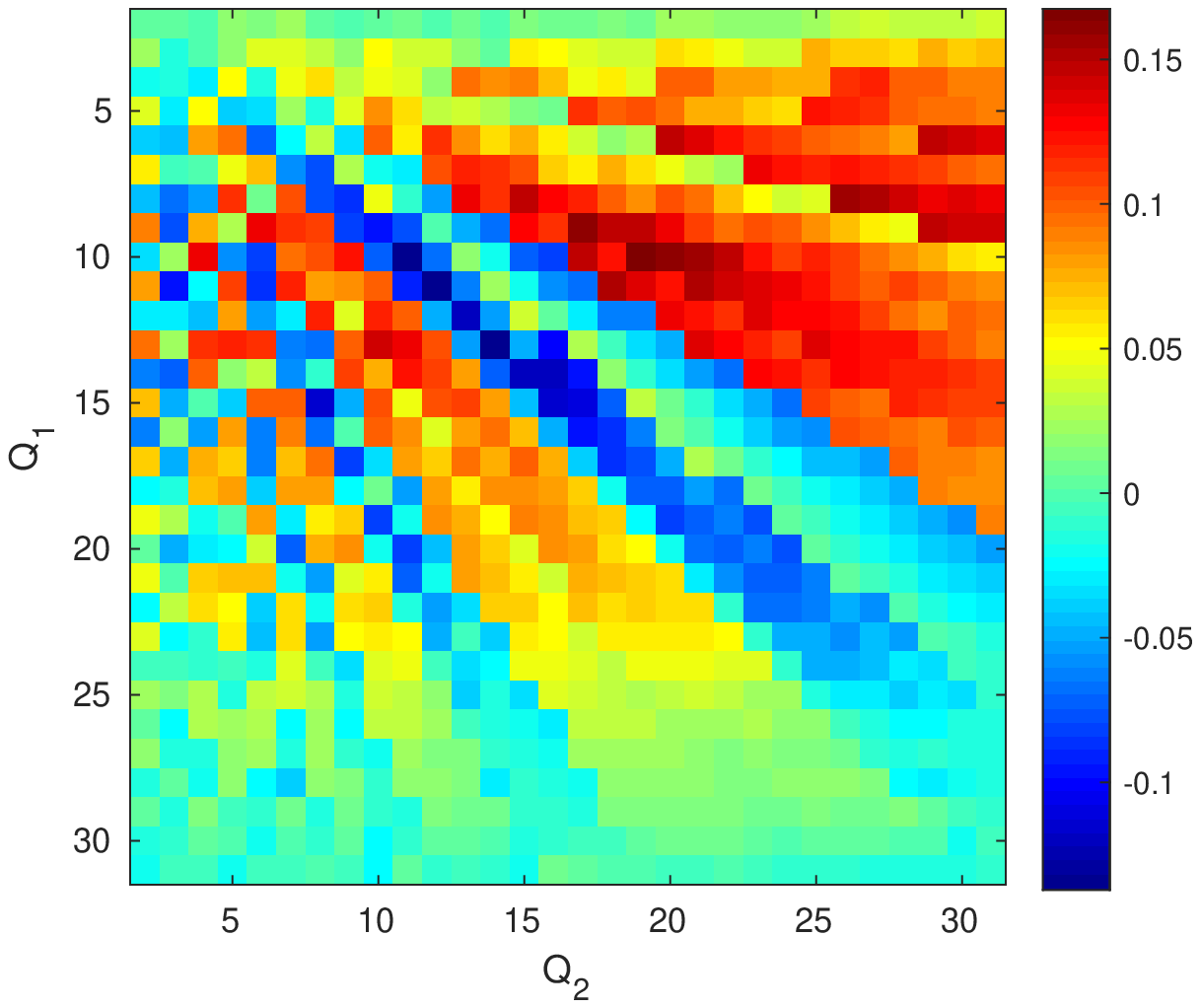}
\centerline{(b1)}
\end{minipage}
\begin{minipage}{0.32\linewidth}
\centering
\includegraphics[width=\linewidth]{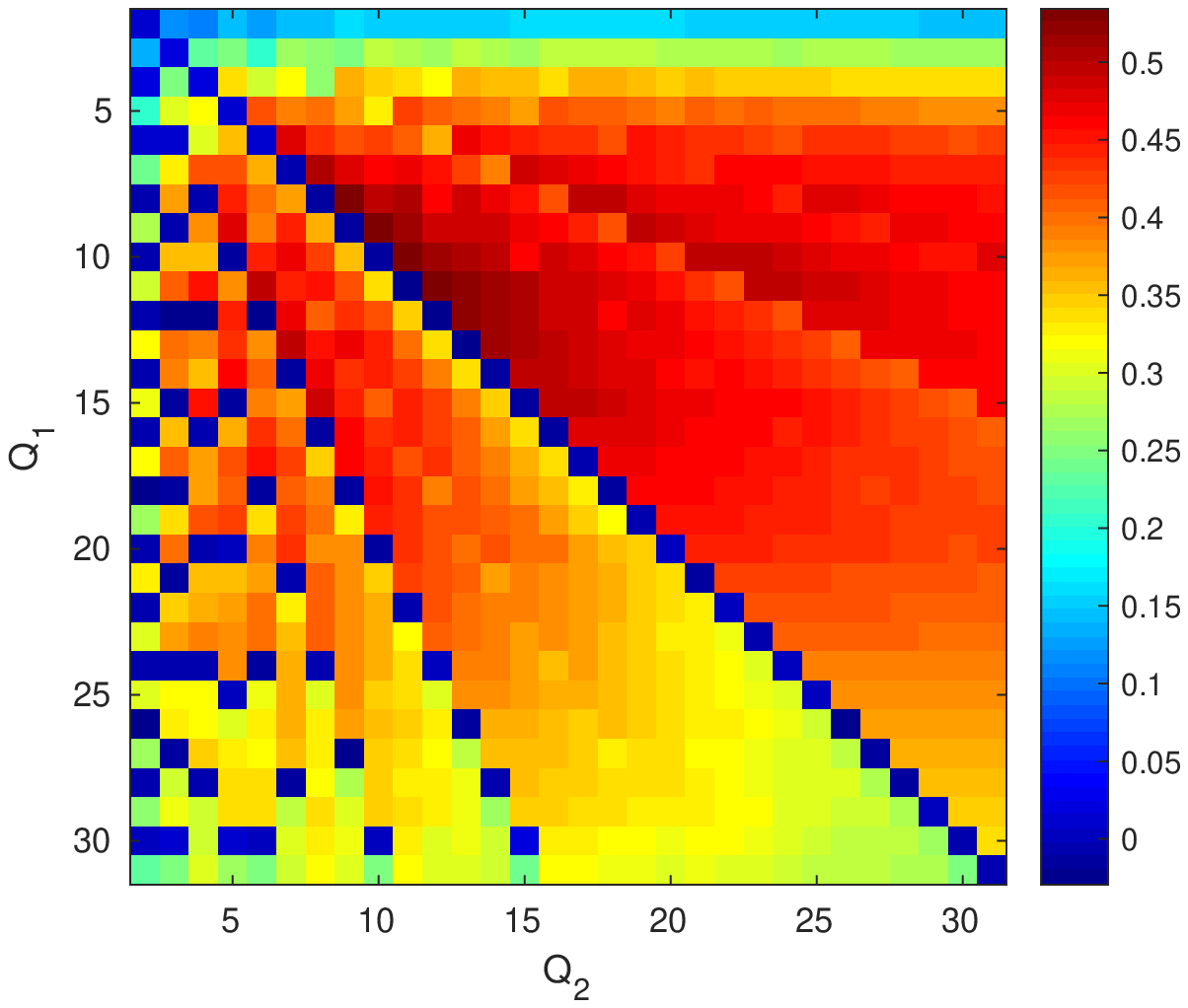}
\centerline{(c1)}
\end{minipage}
\\
\begin{minipage}{0.32\linewidth}
\centering
\includegraphics[width=\linewidth]{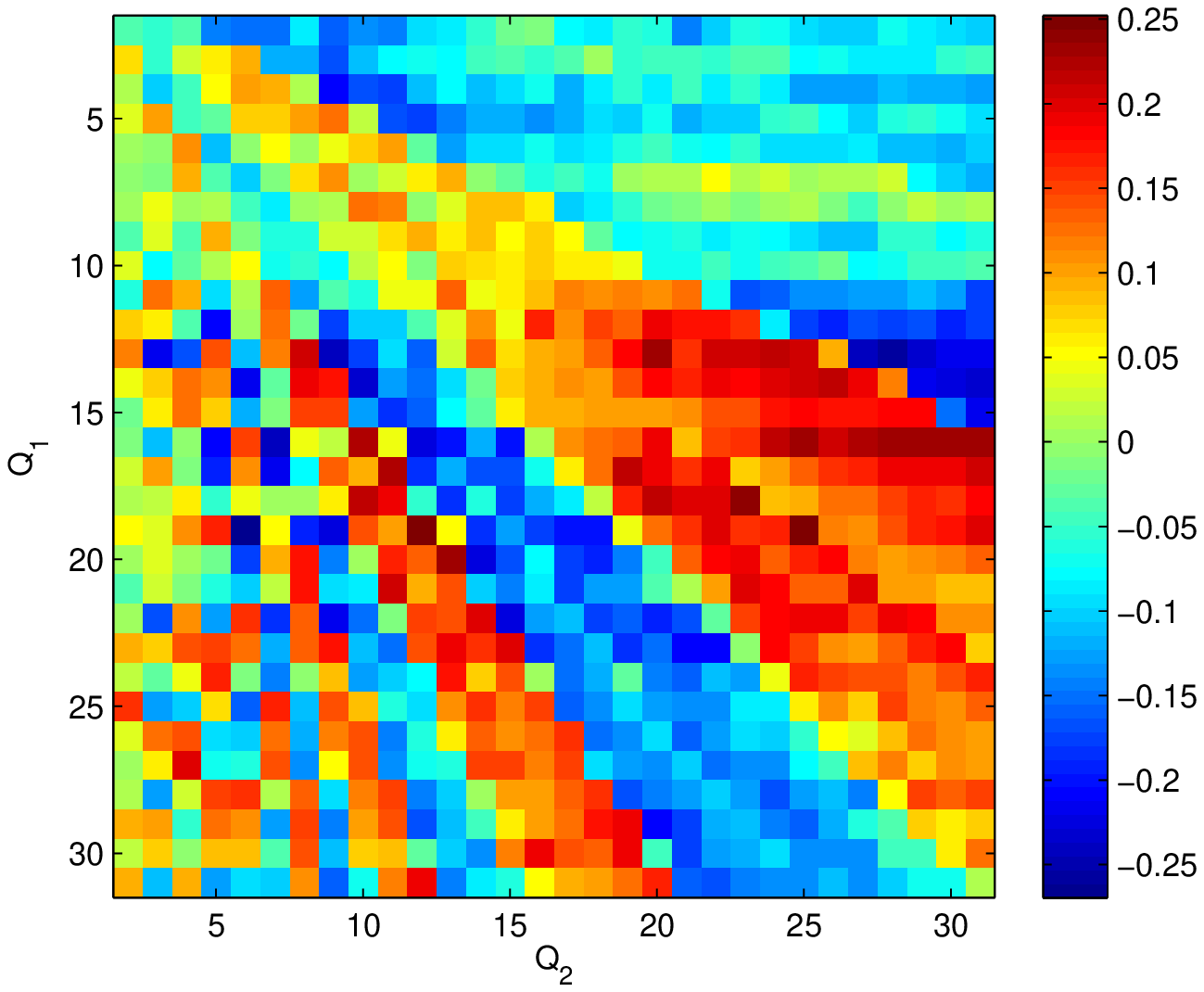}
\centerline{(a2)}
\end{minipage}
\begin{minipage}{0.32\linewidth}
\centering
\includegraphics[width=\linewidth]{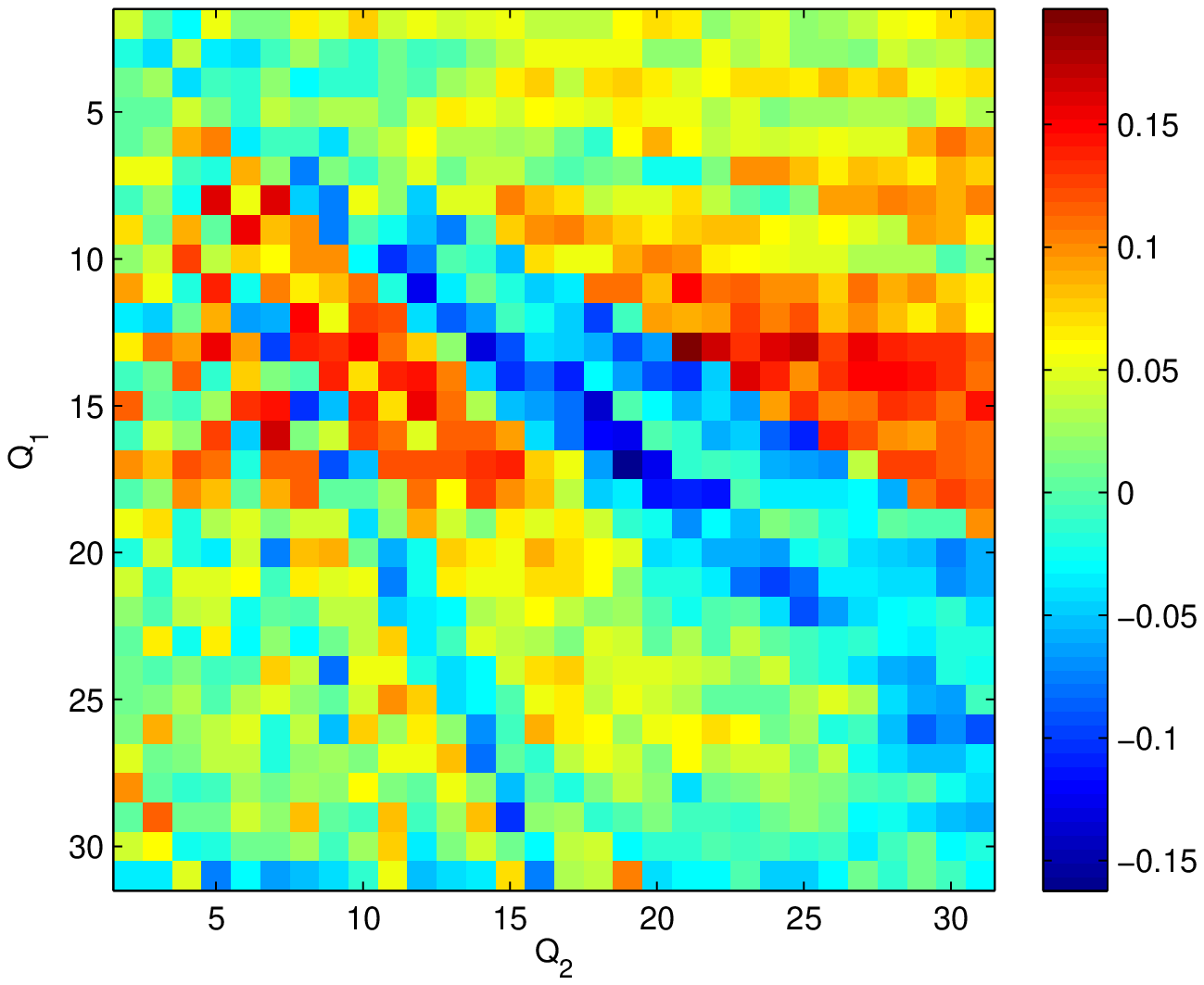}
\centerline{(b2)}
\end{minipage}
\begin{minipage}{0.32\linewidth}
\centering
\includegraphics[width=\linewidth]{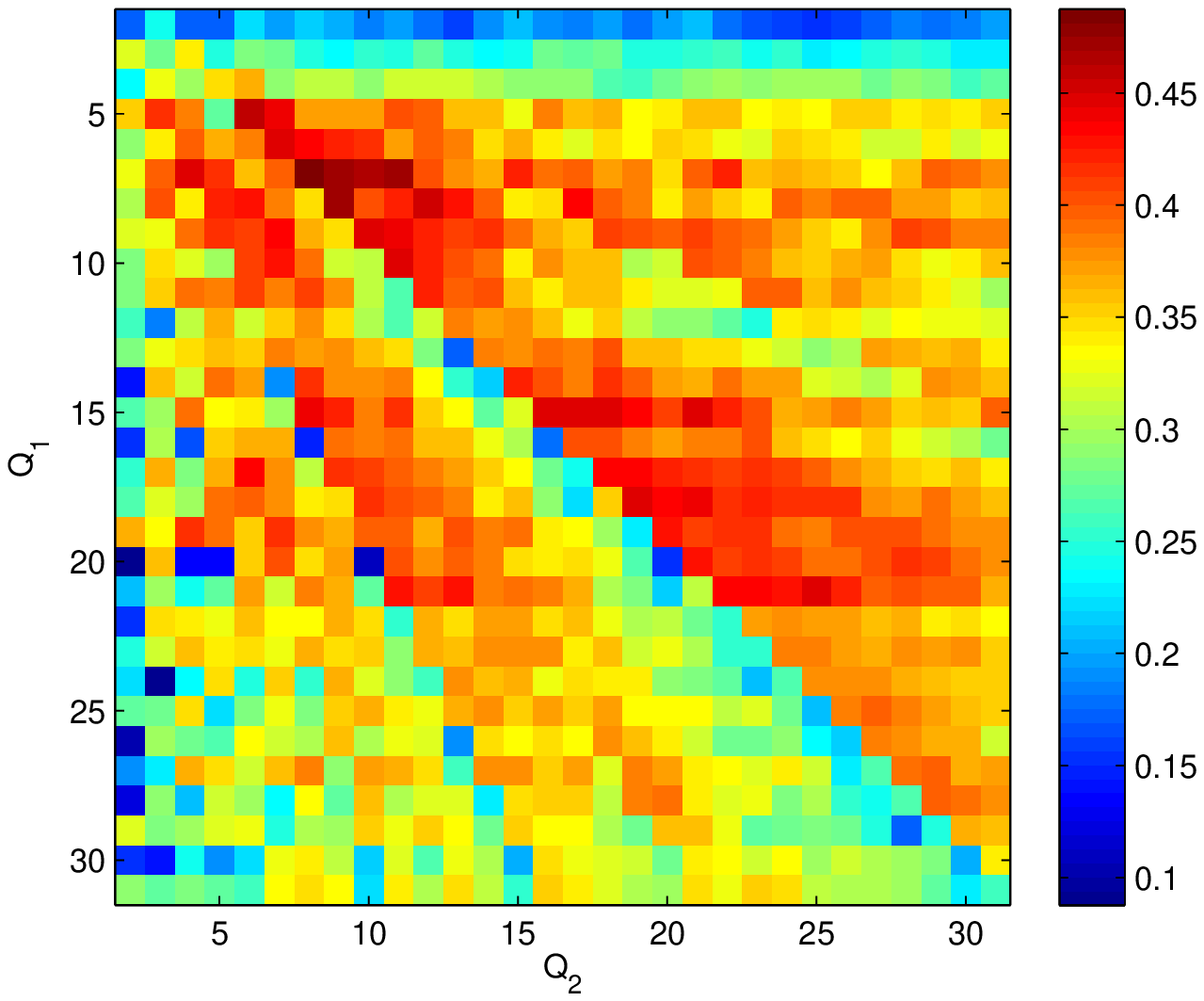}
\centerline{(c2)}
\end{minipage}
\\
\begin{minipage}{0.32\linewidth}
\centering
\includegraphics[width=\linewidth]{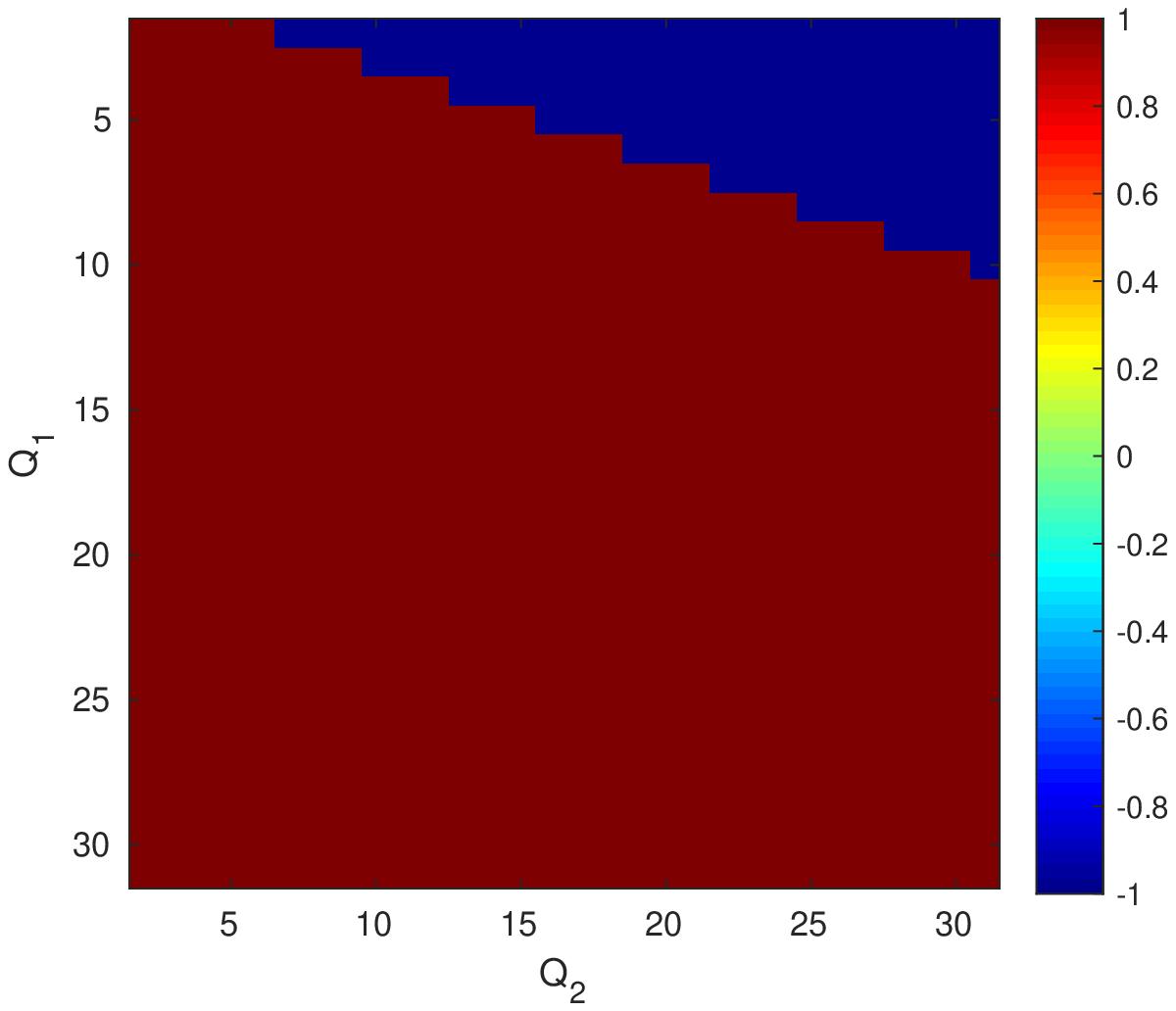}
\centerline{(a3)}
\end{minipage}
\begin{minipage}{0.32\linewidth}
\centering
\includegraphics[width=\linewidth]{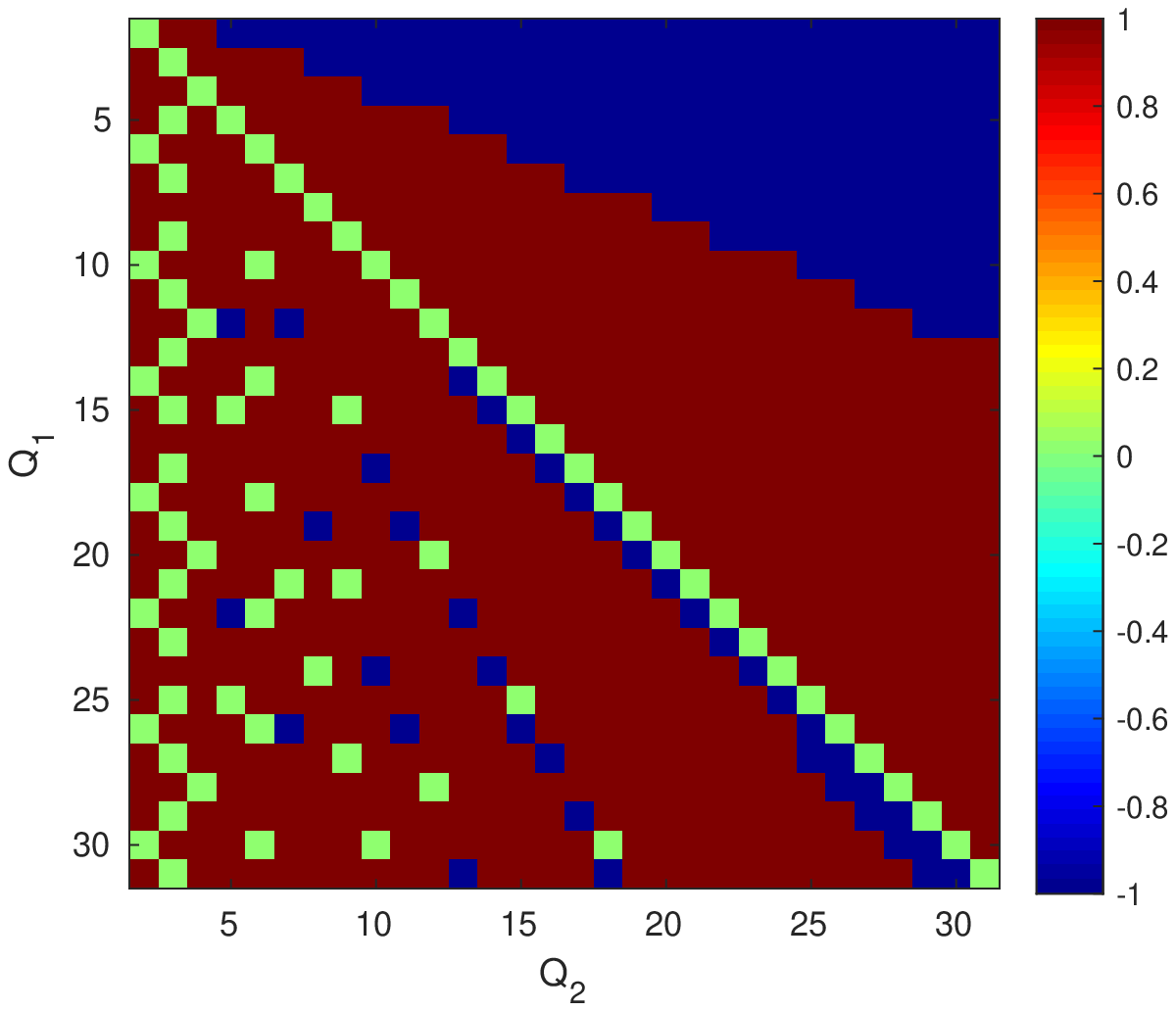}
\centerline{(b3)}
\end{minipage}
\begin{minipage}{0.32\linewidth}
\centering
\includegraphics[width=\linewidth]{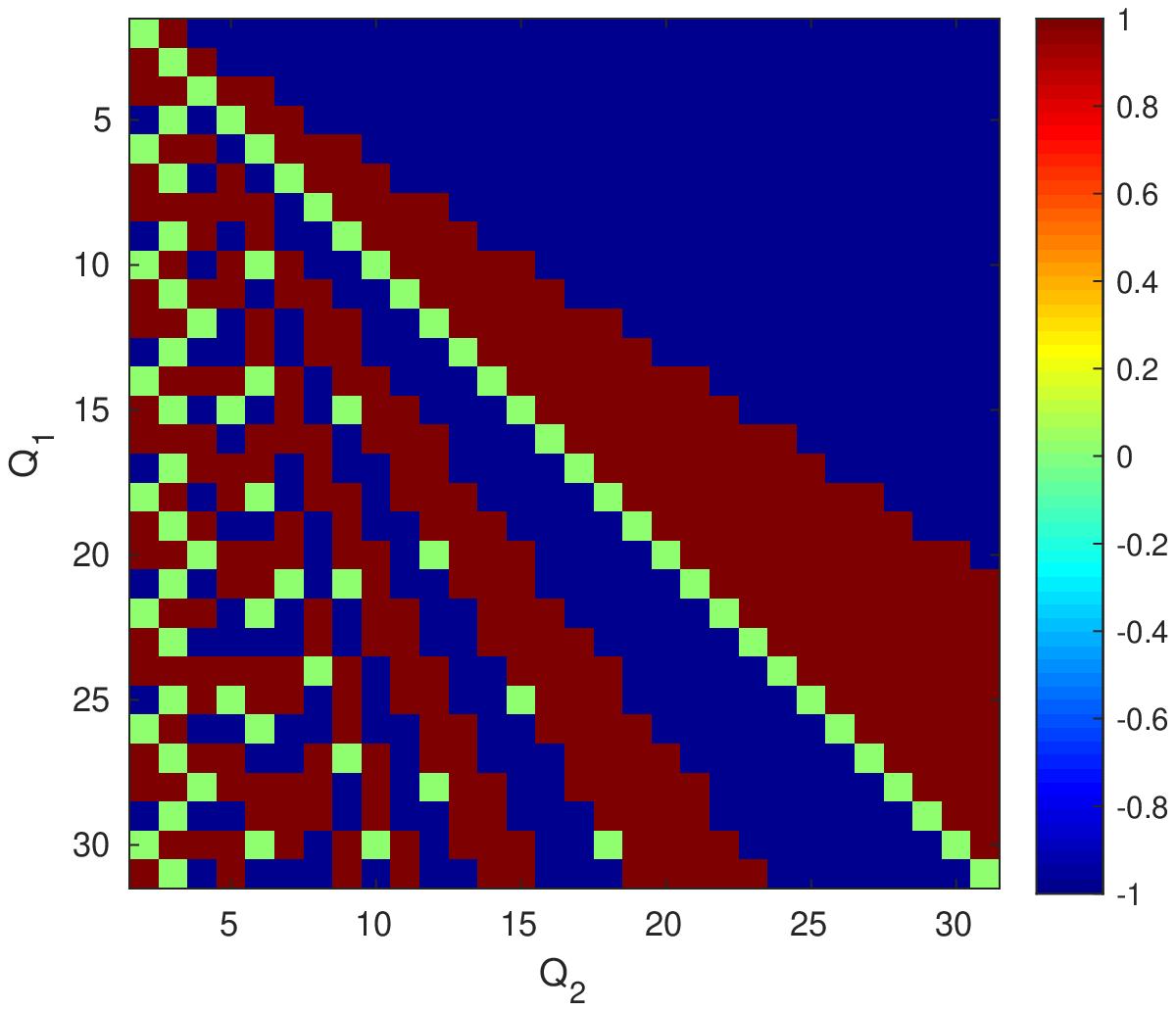}
\centerline{(c3)}
\end{minipage}
\caption{Evolution of $\text{corr}\left(\mathbf{E}_n^{\text{P}_1},\mathbf{E}_{n-1}^{\text{P}_2}\right)$ for a fixed $\alpha_{\text{P}}=2$ and varying $\alpha_{\text{I}}$, $\text{Q}_1$, and $\text{Q}_2$. The upper panels show the obtained results with synthetic signals, while the center panels show the corresponding results for the static video \texttt{akiyo}. In the lower panels, only the sign of the correlation is shown for those samples that have been quantized to the centroid $\Delta_1^{\text{P}}+\frac{\Delta_1^{\text{P}}}{2}$ in the first compression stage. (a1, a2) $\alpha_{\text{I}}=1$, (b1, b2) $\alpha_{\text{I}}=\frac{5}{4}$, (c1, c2) $\alpha_{\text{I}}=2$.}
\label{fig:corr_EP_E3}
\end{figure*}

\item $\text{cov}\left(\mathbf{E}_n^{\text{P}_1},\mathbf{E}_{n-1}^{\text{P}_2}\right)$: here we analyze the correlation between the quantization error $\mathbf{E}_n^{\text{P}_1}$ (whose synthetic version has been detailed in Section~\ref{subsec:intra}) and the one described in the previous point $\mathbf{E}_{n-1}^{\text{P}_2}$. The upper panels of Fig.~\ref{fig:corr_EP_E3} report the obtained values of $\text{corr}\left(\mathbf{E}_n^{\text{P}_1},\mathbf{E}_{n-1}^{\text{P}_2}\right)$ using the described model for a fixed $\alpha_{\text{P}}$ and varying the parameters $\alpha_{\text{I}}$, $\text{Q}_1$, and $\text{Q}_2$. The center panels in Fig.~\ref{fig:corr_EP_E3} show their empirical counterparts, which have been extracted from the low-motion video \texttt{akiyo}. Again, the model remarkably follows the shape of the empirical correlations. In this case, by considering a similar example as in the previous point for $\text{sgn}(d^\prime-d)$, but taking now $d=\Delta_1^{\text{P}}+\frac{\Delta_1^{\text{P}}}{2}$ and reconstructing $d^\prime $ with $\Delta_2^{\text{P}}$ following the rule for P-MBs, one can observe in Fig.~\ref{fig:corr_EP_E3}(c3) that the evolution of $\text{sgn}(d^\prime-d)$ shows a change of sign at $\text{Q}_2>(3/2)\text{Q}_1$, that will determine the limit beyond which the VPF vanishes for $\alpha_{\text{I}}=\alpha_{\text{P}}$.

\end{enumerate}

\begin{figure*}[t]
\centering
\begin{minipage}{0.32\linewidth}
\centering
\includegraphics[width=\linewidth]{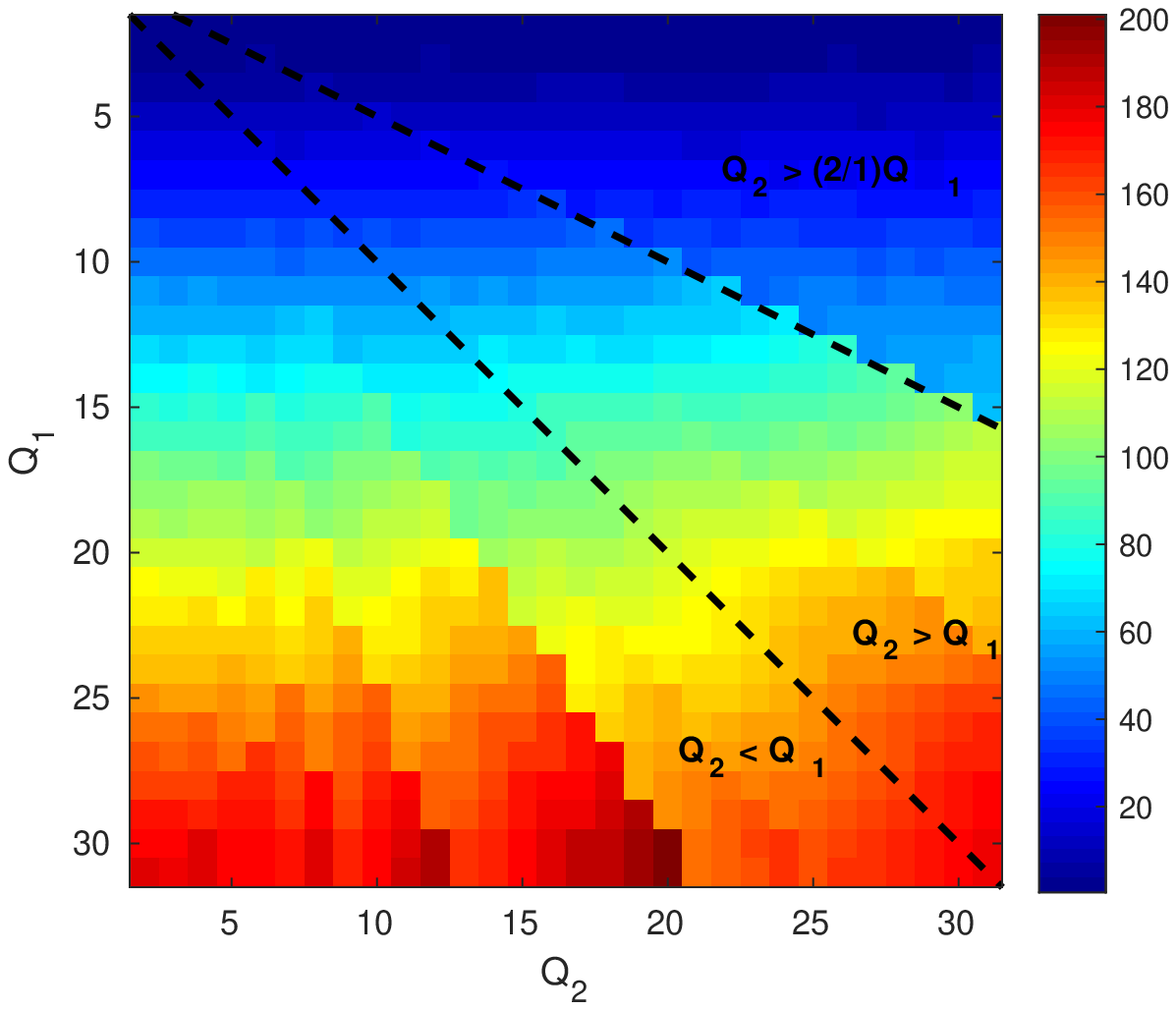}
\centerline{(a1)}
\end{minipage}
\begin{minipage}{0.32\linewidth}
\centering
\includegraphics[width=\linewidth]{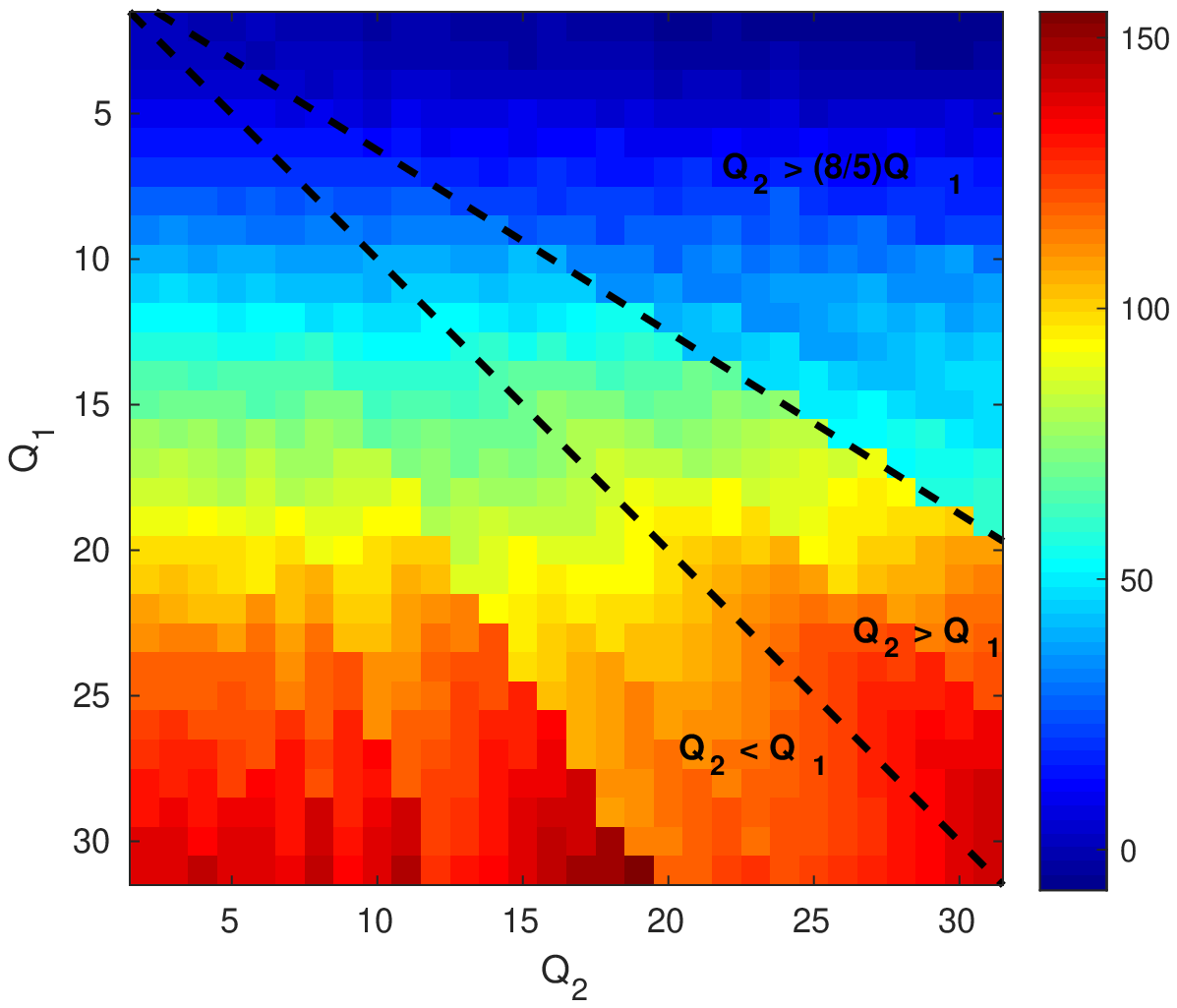}
\centerline{(b1)}
\end{minipage}
\begin{minipage}{0.32\linewidth}
\centering
\includegraphics[width=\linewidth]{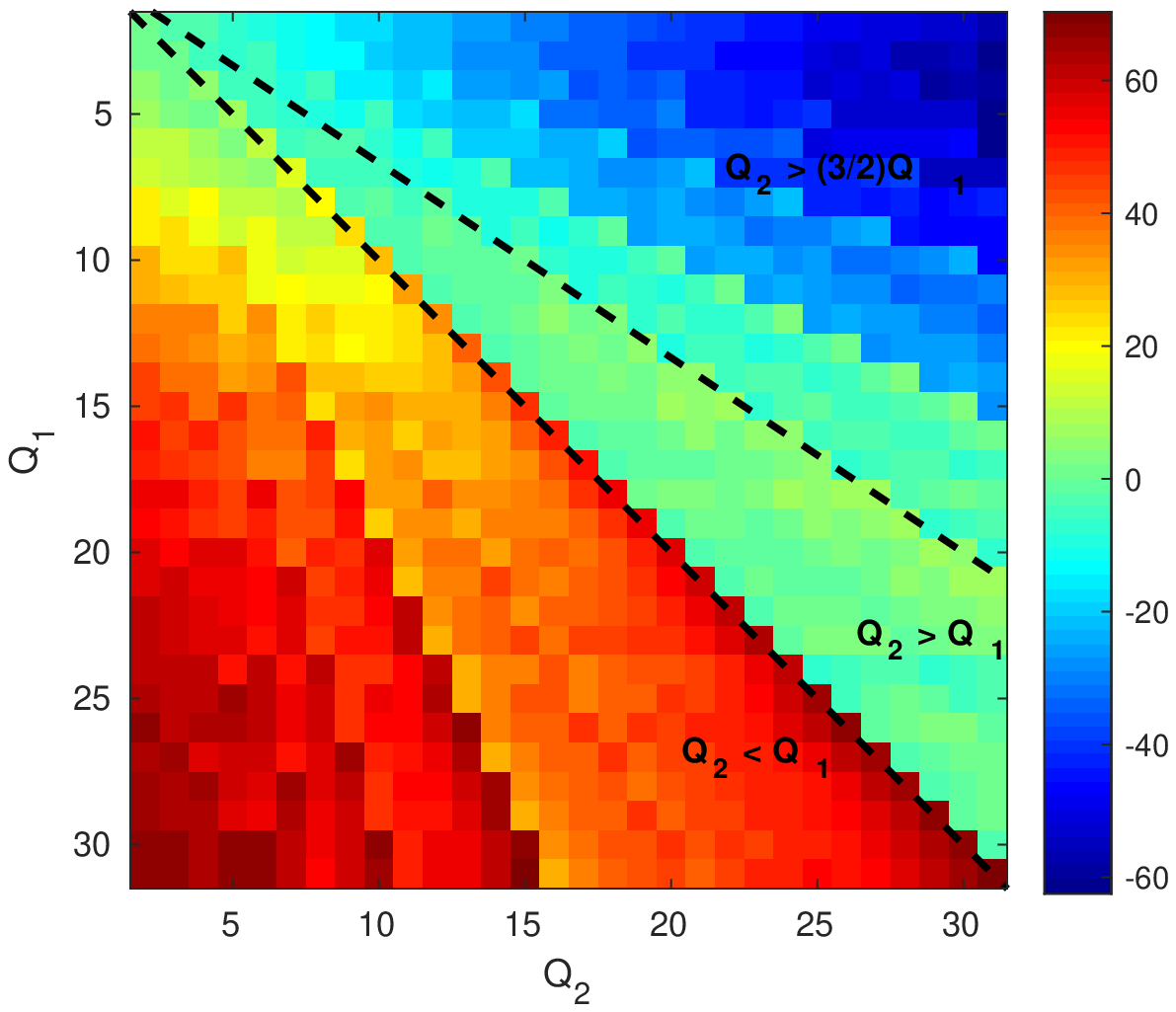}
\centerline{(c1)}
\end{minipage}
\\
\begin{minipage}{0.32\linewidth}
\centering
\includegraphics[width=\linewidth]{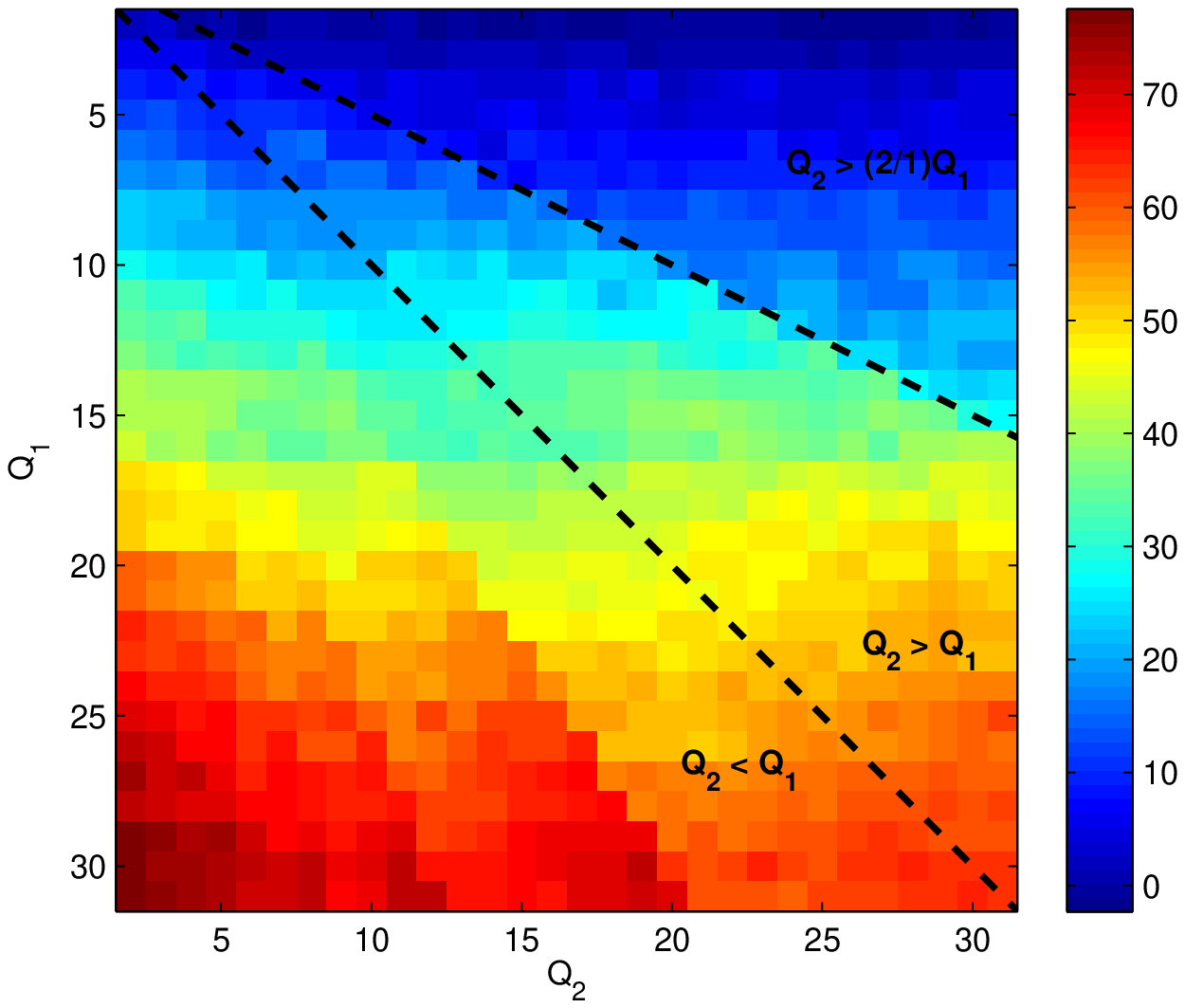}
\centerline{(a2)}
\end{minipage}
\begin{minipage}{0.32\linewidth}
\centering
\includegraphics[width=\linewidth]{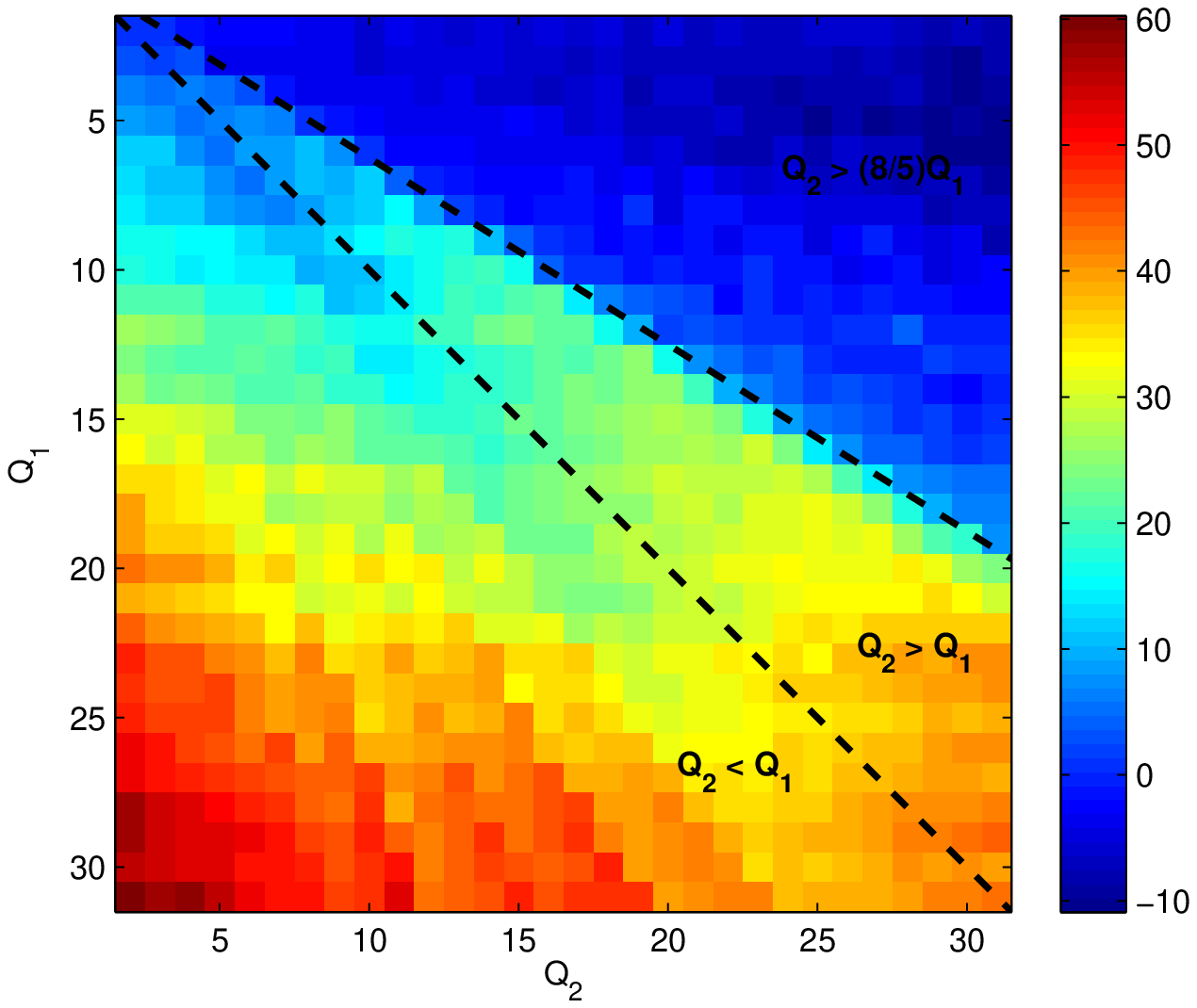}
\centerline{(b2)}
\end{minipage}
\begin{minipage}{0.32\linewidth}
\centering
\includegraphics[width=\linewidth]{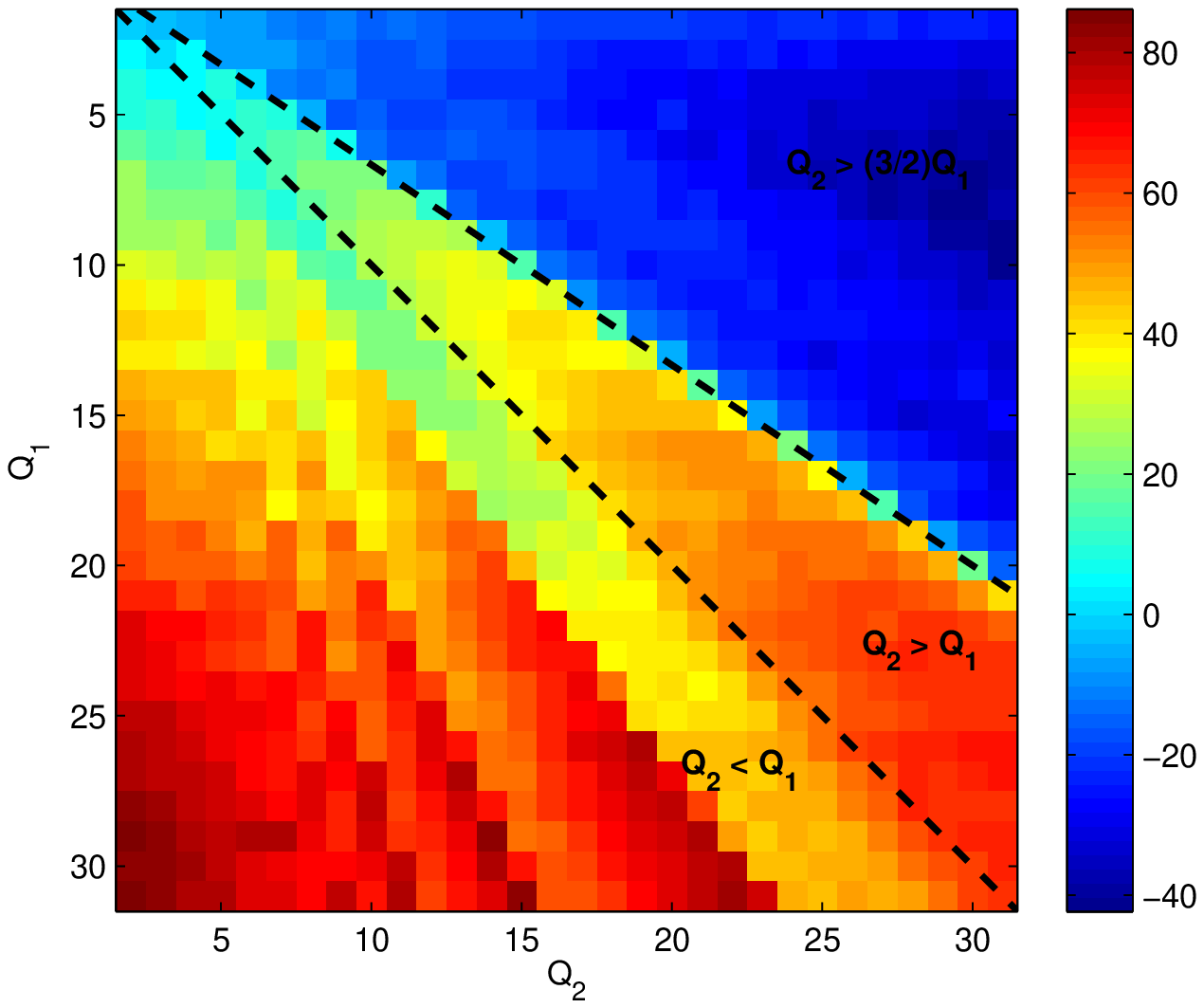}
\centerline{(c2)}
\end{minipage}
\caption{Evolution of $\text{Var}\left(\mathbf{W}_n\right)|_{n\in\mathsf{I}_1}-\text{Var}\left(\mathbf{W}_n\right)|_{n\in\mathsf{P}_1}$ for a fixed $\alpha_{\text{P}}=2$ and varying $\alpha_{\text{I}}$, $\text{Q}_1$, and $\text{Q}_2$. The upper panels show the obtained results with synthetic signals, while the lower panels show the corresponding average difference from 14 real videos in \cite{XIPH}. (a1, a2) $\alpha_{\text{I}}=1$, (b1, b2) $\alpha_{\text{I}}=\frac{5}{4}$, (c1, c2) $\alpha_{\text{I}}=2$.}
\label{fig:diff_W}
\end{figure*}

Once modeled and discussed all the terms in \eqref{eq:diff_sigma_E}, we finally compare in Fig.~\ref{fig:diff_W} the resulting synthetic versions of the difference $\text{Var}\left(\mathbf{W}_n\right)|_{n\in\mathsf{I}_1}-\text{Var}\left(\mathbf{W}_n\right)|_{n\in\mathsf{P}_1}$ for the considered values of $\alpha_{\text{I}}$ (\emph{upper panels}) with the ones empirically obtained after processing 14 videos from \cite{XIPH}\footnote{In particular, we considered the following 14 CIF resolution videos: \texttt{akiyo}, \texttt{bridge-close}, \texttt{bridge-far}, \texttt{container}, \texttt{foreman}, \texttt{hall}, \texttt{highway}, \texttt{mother-daughter}, \texttt{mobile}, \texttt{news}, \texttt{silent}, \texttt{paris}, and \texttt{waterfall}.} (\emph{lower panels}). The synthetic models show a very high degree of similarity with respect to their empirical counterparts, except for the case $\alpha_{\text{I}}\!=\!2$, where the model possibly needs some adjustment. Still, this allows us to predict the strength of the VPF under each particular configuration of $\alpha_{\text{I}}$ and $\alpha_{\text{P}}$, and also to infer from which relation between $\text{Q}_1$ and $\text{Q}_2$ the VPF is more likely to show up or not. In this sense, the limits in the correlation maps previously discussed in the two above points, show the boundary beyond which the difference $\text{Var}\left(\mathbf{W}_n\right)|_{n\in\mathsf{I}_1}-\text{Var}\left(\mathbf{W}_n\right)|_{n\in\mathsf{P}_1}$ drops and, as a consequence, the VPF vanishes. In particular, we have seen that for $\alpha_{\text{I}}\in\{1,\frac{5}{4}\}$, this limit is achieved at $\text{Q}_2>(2/\alpha_{\text{I}})\text{Q}_1$ (see Figs.~\ref{fig:diff_W}(a2)-(b2)), while it moves downward to $\text{Q}_2>(3/2)\text{Q}_1$ for $\alpha_{\text{I}}=\alpha_{\text{P}}$, as can be observed in Fig.~\ref{fig:diff_W}(c2). Given that in all cases, the limit is above $\text{Q}_2>\text{Q}_1$, this explains why the VPF-based approaches are able to satisfactorily work in the challenging video forensic scenarios where the second compression applied is stronger than the first one. The reader is referred to \cite{EUSIPCO} for checking how this theoretical analysis serves to predict the performance of the VPF-based methods.

\section{Conclusions}
\label{sec:conclusions}

In this report we have delved into the analysis of the prediction residue computed in the second stage of an MPEG-2 double compression scheme. The characterization of the quantization process and the different parameters that control the deadzone width of the quantizers have made possible the derivation of a semi-analytic model that through the use of synthetic signals allows us to explain why the VPF shows up and how this footprint behaves depending on the quantization strength applied in each compression stage. One of the most valuable outcomes from the above analysis is the justification of why the approaches that exploit the VPF are able to successfully work on challenging scenarios where the second compression applied is stronger than the first one, while other available techniques typically fail. Nevertheless, the obtained synthetic results are not always consistent with their empirical counterparts, so a review of the semi-analytic model detailed in this report is still necessary, so as to understand what can be causing such inconsistencies.

Furthermore, as pointed out throughout the report, there is room for improving the above analysis, for instance, the model should be extended to encompass other video coding standards (e.g., MPEG-4 and H.264), to address more complex coding settings (e.g., including B-frames, adaptive bitrate controls, etc.), and also to cover more complex type of scenes, such as the dynamic ones. Finally, since the proposed semi-analytic model has proved to be valid, the corresponding closed-form expressions should also be derived in a future work.

\bibliographystyle{IEEEtran}

\bibliography{IEEEabrv,bare_jrnl}

\end{document}